\documentclass[english,12pt, draftclsnofoot, onecolumn]{IEEEtran}
\usepackage[T1]{fontenc}
\usepackage{color}
\usepackage{babel}
\usepackage{float}
\usepackage{amsmath}
\usepackage{amssymb}
\usepackage{graphicx}
\usepackage[unicode=true]
 {hyperref}

\makeatletter

\providecommand{\tabularnewline}{\\}
\floatstyle{ruled}
\newfloat{algorithm}{tbp}{loa}
\providecommand{\algorithmname}{Algorithm}
\floatname{algorithm}{\protect\algorithmname}

\ifCLASSOPTIONcompsoc
\usepackage[caption=false,font=normalsize,labelfont=sf,textfont=sf]{subfig}
\else
\usepackage[caption=false,font=footnotesize]{subfig}
\fi

\usepackage{psfrag}\usepackage{babel}\usepackage{algorithmic}

\usepackage{pgfplots}
\pgfplotsset{width=9cm,compat=1.5}
\usepackage{pgfplotstable}

\makeatother

\begin{document}

\title{Precoder Design with Limited Feedback and Backhauling for Joint Transmission}

\author{Tilak~Rajesh~Lakshmana, Antti~Tölli,~\IEEEmembership{Senior Member,~IEEE,}~\\
Rahul Devassy,~\IEEEmembership{Student Member,~IEEE,} and~Tommy~Svensson,~\IEEEmembership{Senior~Member,~IEEE}%
\thanks{Tilak~Rajesh~Lakshmana, Rahul Devassy, and Tommy~Svensson are with
the Department of Signals and Systems, Chalmers University of Technology,
Gothenburg, Sweden, e-mail: \protect\href{mailto:tilak@chalmers.se}{tilak@chalmers.se},
\protect\href{mailto:devassy@chalmers.se}{devassy@chalmers.se}, \protect\href{mailto:tommy.svensson@chalmers.se}{tommy.svensson@chalmers.se}.%
}%
\thanks{Antti~Tölli is with the Centre for Wireless Communications, University
of Oulu, Oulu, Finland, e-mail: \protect\href{mailto:antti.tolli@ee.oulu.fi }{antti.tolli@ee.oulu.fi }.
\textcolor{red}{}%
}\vspace{-2cm}
}

\IEEEspecialpapernotice{}
\maketitle
\begin{abstract}
A centralized coordinated multipoint downlink joint transmission in
a frequency division duplex system requires channel state information
(CSI) to be fed back from the cell-edge users to their serving BS,
and aggregated at the central coordination node for precoding, so
that interference can be mitigated. The control signals comprising
of CSI and the precoding weights can easily overwhelm the backhaul
resources. Relative thresholding has been proposed to alleviate the
burden; however, this is at the cost of reduction in throughput. In
this paper, we propose utilizing the long term channel statistics
comprising of pathloss and shadow fading in the precoder design to
model the statistical interference for the unknown CSI. In this regard,
a successive second order cone programming (SSOCP) based precoder
for maximizing the weighted sum rate is proposed. The accuracy of
the solution obtained is bounded with the branch and bound technique.
An alternative optimization framework via weighted mean square error
minimization is also derived. Both these approaches provide an efficient
solution close to the optimal, and also achieve efficient backhauling,
in a sense that the precoding weights are generated only for the active
links. For comparison, a stochastic approach based on particle swarm
optimization is also considered.\end{abstract}
\begin{IEEEkeywords}
branch and bound, CoMP, limited backhauling/feedback, MSE, precoding,
SSOCP, weighted sum rate maximization
\end{IEEEkeywords}

\section{Introduction}

\IEEEPARstart{I}{n} cellular coordinated multipoint (CoMP) transmission
systems, the channel state information (CSI) present at the transmitter
plays an important role in harnessing the gains for joint transmission
CoMP. In the downlink, a group of base stations (BSs) coordinate to
coherently serve a group of users being prone to interference \cite{SZ01}-\cite{DHC+12}.
To mitigate interference in a frequency division duplex (FDD) system,
the users need to estimate the CSI based on the downlink pilots from
the BSs, and then feed it back to its cooperating BSs (typically to
the serving BS). In a centralized architecture, the BSs forward the
CSI to a central coordination node, where the CSI from various users
are accumulated to form the aggregated channel matrix which is used
to design a precoder for mitigating interference. In a decentralized
architecture, the users need to share the CSI between the cooperating
BSs to form the precoding weights. Sharing of CSI and the precoding
weights between the BSs and the central coordination node occurs over
the backhaul. This is typically a microwave or an optical fiber link.

\subsection{Previous work \label{sub:Efficient-backhauling-and}}

In an FDD system, the overhead of feeding back the CSI of all the
cooperating BSs from the user could easily overwhelm the wireless
radio interface and the backhaul resources, especially in a centralized
architecture. In this regard, absolute and relative thresholding \cite{PBG+08}
was proposed to limit the CSI feedback. In particular, relative thresholding
is a process in which the users only feedback those links that fall
within a threshold, say 5 dB, relative to its strongest BS. This results
in limited CSI being available for the precoder design. A user centric
clustering is performed in \cite{DY13}, which is similar to the relative
thresholding performed in our work. However, the aim is finding the
optimal tradeoff between total transmit power and sum backhaul capacity
via power minimization. 

In \cite{PBG+11,YG06}, linear precoding is considered, as it provides
a good tradeoff between complexity and performance. With limited CSI,
in \cite{PBG+11} a linear zero forcing (ZF) precoder with suboptimal
power allocation \cite{ZD04} is considered to achieve the backhaul
signaling load reduction based on a physical (PHY) layer precoding
and a medium access control (MAC) layer scheduling approach. Apart
from the PHY and MAC layer approaches, a predefined constrained backhaul
infrastructure can be included in the precoder design as in \cite{MF07}.
The ZF approach requires a well conditioned aggregated channel matrix
at the central coordination node for channel inversion, which cannot
be guaranteed with limited CSI. This poses constraints on how the
users are selected/scheduled, and it makes it harder to achieve efficient
backhauling. In this paper, the term efficient backhauling is used
to denote the case where the number of precoding weights generated
for the active links is equal to the number of CSI coefficients correspondingly
available for the active links at the central coordination node. Note
that for example with the ZF approach, it is possible to generate
the non-zero precoding weights for non-cooperating BSs and require
them to be nulled to achieve efficient backhauling. In a centralized
architecture, if the central coordination node decides the routing
of user data then this will be based on the precoding weights being
generated for only the active links, instead of making all the user
data to be available at all the cooperating BSs. Such an approach
requires efficient backhauling. Another approach to minimize the backhaul
user data transfer is to jointly design the precoder and simultaneously
minimize the user data transfer in the backhaul based on the quality
of service \cite{ZQL13}. In \cite{LBS12}, a stochastic precoder
based on particle swarm optimization (PSO) is designed to achieve
efficient backhauling while taking the limited feedback and limited
backhaul into account. However, with suboptimal power allocation and
with increase in the problem size, the complexity of the algorithm
increases as pointed out in \cite[sec. 3.4]{LBS12}. 

Weighted sum rate maximization is a difficult non-convex problem \cite{SR+11,JWC+12}.
In this regard, different centralized successive convex approximation
(SCA) methods are proposed in \cite{VTT+14}-\cite{HTT+13}. In \cite{THT+12},
a low complexity approximation with faster convergence rate is proposed
for a downlink multicell multiple input single output (MISO) system.
In \cite{HTT+13}, a different approximation is used for robust precoding
with uncertainty in CSI at the transmitter. The precoders can also
be designed via the mean square error (MSE) approach. In \cite{SR+11,CAC+08},
it was shown that minimizing the weighted sum mean square error (MSE)
is equivalent to the weighted sum rate maximization, where the precoder,
receive weight and the receiver MSE weights are alternately optimized.
Whenever a central coordination node is not available, then \cite{SR+11,KA+13}
can be used to implement the precoder in a decentralized fashion.
Signaling strategies are considered in \cite{KA+13}, and also under
imperfect channel conditions \cite{FF13} extending the result from
\cite{CAC+08}. Also, in \cite{AS14}, a generalized mean square error
criterion is used to arrive at a robust linear precoding solution
that can handle backhaul constraints with CSI uncertainty. Similar
to weighted sum rate maximization, a cross layer queue deviation minimization
is considered in \cite{VTT+14} where the queue states act as weights
for the sum rate maximization, with a different approximation of the
signal to interference plus noise (SINR) constraint.

\subsection{Contributions\label{sub:Contributions}}

In this work, we focus on the design of the precoder in a centralized
FDD system with the objective of maximizing the weighted sum rate
of the users with perfect but limited CSI feedback, and also under
limited backhauling. We use the algorithms developed for the full
CSI case, but now we incorporate the limited CSI and the statistical
model of interference. In this regard, we propose a conservative precoder
design for any combination of user centric clustering with per-antenna
power constraint, where the long term channel statistics is incorporated
into the optimization problem for the missing links. Here, we model
the statistical interference for the unknown CSI as the long term
channel statistics in the interference terms for the user. The model
is pessimistic in nature, as the Cauchy-Schwarz inequality is applied
on the unknown parts that were separated from the total interference.
The long term channel statistics is also used for making the routing
decisions for the user data in the backhaul as noted in \cite{KG12}. 

In our work, we effectively solve the problem of designing a PHY layer
precoder with limited information based on the approach in \cite{VTT+14}-\cite{HTT+13},
where we extend the SCA framework to cope with incomplete CSI at the
transmitter. Here we consider joint transmission CoMP while \cite{VTT+14}-\cite{HTT+13}
focused on coordinated beamforming. In this regard, we incorporate
the pessimistic interference model based on the long term channel
statistics (pathloss and shadow fading) into the problem formulation.
Alternatively we also include this statistical interference model
in the minimization of the weighted MSE \cite{CAC+08}-\cite{KA+13}.
We also use the long term channel statistics to determine the CSI
feedback threshold. Our proposed pessimistic statistical interference
modeling is different compared to the previous work \cite{PBG+08,PBG+11,LBS12,DSI+12}
where the unavailable CSI are modeled as zeros. The availability of
the long term statistics at the coordination node is a valid assumption,
as they are available in the existing cellular standards, where the
users feedback the received signal strength, more popularly referred
to as the received signal strength indicator (RSSI). For example,
this is required during handover procedures. In our setup, we consider
relative thresholding, a variant of \cite{PBG+08}, based on this
average signal strength at the user. 

The main contributions of this work are listed as follows:
\begin{itemize}
\item The long term channel statistics (pathloss and shadow fading), based
on relative thresholding, are modeled in the precoder design as part
of the pessimistic statistical interference in the SINR ratio. 
\item We efficiently solve the precoder design problem with the limited
feedback and limited backhauling, using a successive second order
cone programming (SSOCP). Also, we solve for the case when the long
term channel statistics is considered as part of the SOC constraint,
instead of neither treating them as zeros nor naively replacing the
zeros with this side information. We also achieve efficient backhauling,
where the precoding weights are generated only for those links whose
CSI was reported.
\item As an alternative to the SSOCP, we reformulate the problem via weighted
MSE criterion similar to \cite{SR+11,CAC+08}, with the use of the
proposed long term channel statistics in the variance of the received
signal. The results show that it achieves the same weighted sum rate
as that of the proposed SSOCP on average. The MSE reformulation requires
a higher number of iterations than SSOCP to converge but each sub-problem
is simple to solve.
\item We characterize the performance of the proposed precoder design using
numerical bounds with a variant of the branch and bound technique
\cite{JWC+12}. The proposed iterative SSOCP algorithm is very close
to the optimal provided by the branch and bound method.
\item We numerically compare the performance of the proposed iterative algorithm
to an existing stochastic algorithm under limited feedback and limited
backhauling. In particular we consider PSO, as the overhead of book
keeping of variables is far simpler compared to other stochastic algorithms
such as ant colony optimization or evolutionary algorithms. It was
found that the performance of the PSO is inferior, especially when
the problem size is increased.
\end{itemize}
The paper is organized as follows: the system model is introduced
in Section~\ref{sec:System-Model}. In Section~\ref{sec:Precoder-design},
the precoder design based on SSOCP and MSE are derived. A brief description
of PSO is presented, and this section concludes with the branch and
bound technique used to bound the performance of SSOCP. Using the
derived precoders, the simulation results are presented in Section~\ref{sec:Simulations},
in terms of the effect of threshold, cell-edge signal to noise ratio
(SNR), BS antennas and the SSOCP bounds. Finally Section~\ref{sec:Conclusions}
concludes the contribution of the paper.

Notation: A scalar variable is denoted as $x$ while $X$ denotes
a scalar constant. A vector and a matrix are denoted as $\mathbf{x}$
and $\mathbf{X}$, respectively. A set is denoted in calligraphic
font as $\mathcal{X}$ and the cardinality of the set is $\left|\mathcal{X}\right|$.
The elements of set $\mathcal{X}$ not in set $\mathcal{Y}$ is denoted
as $\mathcal{X}\backslash\mathcal{Y}$. The absolute value of $x\mathbb{\in C}$
is denoted as $|x|$ while the $p-$norm of a vector is denoted as
$||\cdot||_{p}$. The transpose and conjugate transpose of a vector
$\mathbf{x}$ is denoted as $\mathbf{x}^{T}$ and $\mathbf{x}^{H}$,
respectively. The expectation operation on the random variable $X$
is denoted as $E_{X}\left[X\right]$.

\section{System Model\label{sec:System-Model}}

Consider a homogenous network cluster consisting of $\left|\mathcal{B}\right|$
BSs, each with $N_{\text{T}}$ antennas. The BSs are coordinated to
serve $\left|\mathcal{U}\right|$ single antenna cell-edge users.
The signal received by the $u$th user is $y_{u}$, and it consists
of the desired signal and intracluster interference
\begin{equation}
y_{u}=\underset{b\in\mathcal{B}_{u}}{\sum}\mathbf{h}_{b,u}\mathbf{w}_{b,u}x_{u}+\underset{i\neq u}{\sum}\underset{b\in\mathcal{B}_{i}}{\sum}\mathbf{h}_{b,u}\mathbf{w}_{b,i}x_{i}+n_{u},\label{eq:received sig}
\end{equation}
where $\mathcal{B}_{u}$ is the set of BSs from which the $u$th user
is served. In this model, the intercluster interference is considered
to be negligible for the cell-edge users located at the cluster center,
and therefore it is not accounted in \eqref{eq:received sig}. The
channel experienced by the $u$th user from $b$th BS with $N_{\text{T}}$
antennas is $\mathbf{h}_{b,u}\in\mathbb{C}^{1\times N_{\text{T}}}$.
The precoding weight for the $u$th user with normalized data $x_{u}$
from the $b$th BS with $N_{\text{T}}$ antennas is $\mathbf{w}_{b,u}\in\mathbb{C}^{N_{\text{T}}\times1}$,
such that $\mathbf{w}_{b,u}=[w_{b,u}^{(1)},w_{b,u}^{(2)},\ldots,w_{b,u}^{(k)},\ldots,w_{b,u}^{(N_{\text{T}})}]^{T}$
where $w_{b,u}^{(k)}$ is the precoding weight on the $k$th antenna
of the $b$th BS for the $u$th user, and $n_{u}$ is the receiver
noise at $u$th user with power $N_{0}$. 

To incorporate the long term channel statistics, let us first consider
the SINR evaluated at the central coordination node for the $u$th
user as\raggedbottom
\begin{eqnarray}
\tilde{\gamma}_{u} & = & \frac{\left|\underset{b\in\mathcal{B}_{u}}{\sum}\mathbf{h}_{b,u}\mathbf{w}_{b,u}\right|^{2}}{\underset{i\neq u}{\sum}\left|\underset{b\in\mathcal{B}_{i}}{\sum}\mathbf{h}_{b,u}\mathbf{w}_{b,i}\right|^{2}+N_{0}}\nonumber \\
 & = & \frac{\left|\underset{b\in\mathcal{B}_{u}}{\sum}\mathbf{h}_{b,u}\mathbf{w}_{b,u}\right|^{2}}{\underset{i\neq u}{\sum}\left\{ \left|\underset{b\in\mathcal{B}_{i}\cap\mathcal{B}_{u}}{\sum}\mathbf{h}_{b,u}\mathbf{w}_{b,i}+\underset{b\in\mathcal{B}_{i}\backslash\mathcal{B}_{u}}{\sum}\mathbf{\overline{h}}_{b,u}\mathbf{w}_{b,i}\right|^{2}\right\} +N_{0}}\label{eq:SINR_Int}
\end{eqnarray}
where the interference terms in the denominator of \eqref{eq:SINR_Int}
are split based on relative thresholding, i.e., the set $\mathcal{B}_{i}\cap\mathcal{B}_{u}$
denotes the set of BSs that are involved in serving both the $u$th
and the $i$th user, as the CSI $\mathbf{h}_{b,u}$ falls within the
relative threshold window. However, those links that fall outside
this threshold constitute the term $\mathbf{\overline{h}}_{b,u}$
where $\mathcal{B}_{i}\backslash\mathcal{B}_{u}$ is the set of BSs
serving the $i$th user but not the $u$th user. The given set $\mathcal{B}_{u}$
is defined by the relative thresholding algorithm based on the long
term channel statistics as summarized in Algorithm~\ref{alg:Relative-thresholding-performed}.
To achieve the condition of efficient backhauling, the precoding weights
are generated only for those links for which the users have fed back
the CSI. 
\begin{algorithm}[H]
\begin{centering}
\protect\caption{\label{alg:Relative-thresholding-performed}Relative thresholding
performed at the user based on the long term channel statistics (pathloss
and shadow fading)}

\par\end{centering}

\small{

\begin{algorithmic}[1]

\STATE Set the feedback threshold, $T\left(=3\,\mbox{dB, for example}\right)$

\FOR { $\forall u\in\mathcal{U}$ }

\STATE Perform channel measurements of the BSs, $\mathcal{B}$

\STATE $c=\underset{b\in\mathcal{B}}{\mbox{max}}\left(E\left[||\mathbf{h}_{b,u}||_{2}^{2}\right]\right)$

\FOR { $\forall b\in\mathcal{B}$ }

\IF{ $\left(c_{\text{dB}}-\left[E\left[||\mathbf{h}_{b,u}||_{2}^{2}\right]\right]_{\text{dB}}\right)\leq T$
}

\STATE Include $b$ in the set $\mathcal{B}_{u}$

\ENDIF

\ENDFOR

\STATE The $u$th user feeds back the CSI of the set of BSs in $\mathcal{B}_{u}$

\ENDFOR

\end{algorithmic}

}
\end{algorithm}

Now we define a new SINR, $\overline{\gamma}_{u}$ in \eqref{eq:StartNewSINR},
where we replace the unknown channel coefficients with an expectation
as in \eqref{eq:StartNewSINR}-\eqref{eq:SINR_PL}.\allowdisplaybreaks
\begin{figure*}[tbh]
\begin{subequations}
\begin{eqnarray}
\overline{\gamma}_{u} & \negthickspace\negthickspace\negthickspace= & \negthickspace\negthickspace\negthickspace\frac{\left|\underset{b\in\mathcal{B}_{u}}{\sum}\mathbf{h}_{b,u}\mathbf{w}_{b,u}\right|^{2}}{E_{\overline{h}}\underset{i\neq u}{\sum}\left\{ \left|\underset{b\in\mathcal{B}_{i}\cap\mathcal{B}_{u}}{\sum}\mathbf{h}_{b,u}\mathbf{w}_{b,i}+\underset{b\in\mathcal{B}_{i}\backslash\mathcal{B}_{u}}{\sum}\overline{\mathbf{h}}_{b,u}\mathbf{w}_{b,i}\right|^{2}\right\} +N_{0}}\label{eq:StartNewSINR}\\
 & \negthickspace\negthickspace\negthickspace= & \negthickspace\negthickspace\negthickspace\frac{\left|\underset{b\in\mathcal{B}_{u}}{\sum}\mathbf{h}_{b,u}\mathbf{w}_{b,u}\right|^{2}}{{\scriptstyle \underset{i\neq u}{\sum}\left\{ \left|\underset{b\in\mathcal{B}_{i}\cap\mathcal{B}_{u}}{\sum}\mathbf{h}_{b,u}\mathbf{w}_{b,i}\right|^{2}+E_{\overline{h}}\left|\underset{b\in\mathcal{B}_{i}\backslash\mathcal{B}_{u}}{\sum}\overline{\mathbf{h}}_{b,u}\mathbf{w}_{b,i}\right|^{2}+2\Re\left\{ \left(\underset{b\in\mathcal{B}_{i}\cap\mathcal{B}_{u}}{\sum}\mathbf{h}_{b,u}\mathbf{w}_{b,i}\right)^{H}\underset{b\in\mathcal{B}_{i}\backslash\mathcal{B}_{u}}{\sum}E_{\overline{h}}\left[\overline{\mathbf{h}}_{b,u}\right]\mathbf{w}_{b,i}\right\} \right\} +N_{0}}}\label{eq:ExpectedValueOfhBar}\\
 & \negthickspace\negthickspace\negthickspace= & \negthickspace\negthickspace\negthickspace\frac{\left|\underset{b\in\mathcal{B}_{u}}{\sum}\mathbf{h}_{b,u}\mathbf{w}_{b,u}\right|^{2}}{\underset{i\neq u}{\sum}\left\{ \left|\underset{b\in\mathcal{B}_{i}\cap\mathcal{B}_{u}}{\sum}\mathbf{h}_{b,u}\mathbf{w}_{b,i}\right|^{2}+E_{\overline{h}}\left|\underset{b\in\mathcal{B}_{i}\backslash\mathcal{B}_{u}}{\sum}\overline{\mathbf{h}}_{b,u}\mathbf{w}_{b,i}\right|^{2}\right\} +N_{0}}\\
 & \negthickspace\negthickspace\negthickspace\geq & \negthickspace\negthickspace\negthickspace\frac{\left|\underset{b\in\mathcal{B}_{u}}{\sum}\mathbf{h}_{b,u}\mathbf{w}_{b,u}\right|^{2}}{\underset{i\neq u}{\sum}\left\{ \left|\underset{b\in\mathcal{B}_{i}\cap\mathcal{B}_{u}}{\sum}\mathbf{h}_{b,u}\mathbf{w}_{b,i}\right|^{2}+|\mathcal{B}_{i}\backslash\mathcal{B}_{u}|\underset{b\in\mathcal{B}_{i}\backslash\mathcal{B}_{u}}{\sum}E_{\overline{h}}\left|\overline{\mathbf{h}}_{b,u}\mathbf{w}_{b,i}\right|^{2}\right\} +N_{0}}\label{eq:CauchySchwartz}\\
 & \negthickspace\negthickspace\negthickspace= & \negthickspace\negthickspace\negthickspace\frac{\left|\underset{b\in\mathcal{B}_{u}}{\sum}\mathbf{h}_{b,u}\mathbf{w}_{b,u}\right|^{2}}{\underset{i\neq u}{\sum}\left\{ \left|\underset{b\in\mathcal{B}_{i}\cap\mathcal{B}_{u}}{\sum}\mathbf{h}_{b,u}\mathbf{w}_{b,i}\right|^{2}+|\mathcal{B}_{i}\backslash\mathcal{B}_{u}|\underset{b\in\mathcal{B}_{i}\backslash\mathcal{B}_{u}}{\sum}\lambda_{b,u}^{2}||\mathbf{w}_{b,i}||_{2}^{2}\right\} +N_{0}}\triangleq\gamma_{u}.\label{eq:SINR_PL}
\end{eqnarray}

\end{subequations}
\end{figure*}
 We obtain \eqref{eq:ExpectedValueOfhBar} by expanding the terms
as $|a+b|^{2}=|a|^{2}+|b|^{2}+ab^{H}+ba^{H}=|a|^{2}+|b|^{2}+2\Re\left\{ ab^{H}\right\} $and
taking the expectation inside. Here the hermitian operator is degenerated
to a scalar case.  Here we focus on $E_{\overline{h}}\left[\overline{\mathbf{h}}_{b,u}\right]$,
where $\mathbf{\overline{h}}_{b,u}$ consists of the three random
variables, the pathloss, $l$, the shadow fading, $s_{b,u}\sim\text{ln}\mathcal{N}(0,\sigma_{\text{SF}}^{2})$,
and the small scale fading on the $k$th antenna is $f_{b,u}^{\left(k\right)}\sim\mathcal{CN}\left(0,1\right)$
and $\mathbf{f}_{b,u}\in\mathbb{C}^{1\times N_{\text{T}}}=\left[f_{b,u}^{\left(1\right)},f_{b,u}^{\left(2\right)},\cdots,f_{b,u}^{\left(N_{\text{T}}\right)}\right]$.
The large scale fading and the small scale fading are independent
random variables, and $E_{f}\left[f_{b,u}\right]=0$, therefore we
have $E_{\overline{h}}\left[\overline{\mathbf{h}}_{b,u}\right]=E_{l,s,f}\left[l_{b,u}s_{b,u}\mathbf{f}_{b,u}\right]=E_{l,s}\left[l_{b,u}s_{b,u}\right]E_{r}\left[\mathbf{f}_{b,u}\right]=\mathbf{0}_{N_{\text{T}}}$.
The inequality in \eqref{eq:CauchySchwartz} is obtained using the
Cauchy-Schwarz inequality $\left|\overset{N}{\underset{j=1}{\sum}}a_{j}b_{j}^{*}\right|^{2}\leq\overset{N}{\underset{j=1}{\sum}}\left|a_{j}\right|^{2}\overset{N}{\underset{j=1}{\sum}}\left|b_{j}\right|^{2}$,
where $a_{j}=\overline{\mathbf{h}}_{j,u}\mathbf{w}_{j,i}$ and $b_{j}=1,\forall j$.
Finally, we obtain \eqref{eq:SINR_PL} as follows, $E_{\overline{h}}\left|\overline{\mathbf{h}}_{b,u}\mathbf{w}_{b,i}\right|^{2}=E_{\overline{h}}\left[\mathbf{w}_{b,i}^{H}\mathbf{\overline{h}}_{b,u}^{H}\overline{\mathbf{h}}_{b,u}\mathbf{w}_{b,i}\right]=\mathbf{w}_{b,i}^{H}E_{\overline{h}}\left[\overline{\mathbf{h}}_{b,u}^{H}\overline{\mathbf{h}}_{b,u}\right]\mathbf{w}_{b,i}=\lambda_{b,u}^{2}||\mathbf{w}_{b,i}||_{2}^{2}$,
where $\lambda_{b,u}^{2}$ is the long term channel statistics of
$\mathbf{\overline{h}}_{b,u}$, $E_{\overline{h}}\left[\mathbf{\overline{h}}_{b,u}^{H}\mathbf{\overline{h}}_{b,u}\right]=\lambda_{b,u}^{2}\mathbf{I}_{N_{\text{T}}}$.
We assume that the RSSI is reported by the user and the transmit antennas
are uncorrelated. For correlated channels, covariance matrices $E_{\overline{h}}\left[\mathbf{\overline{h}}_{b,u}^{H}\mathbf{\overline{h}}_{b,u}\right]=\mathbf{C}$
can be incorporated in the problem formulation. Finally, the weighted
sum rate maximization of $\left|\mathcal{U}\right|$ users is evaluated
as
\begin{equation}
R_{\mbox{tot}}=\underset{u}{\sum}\alpha_{u}\mbox{log}_{2}\left(1+\gamma_{u}\right)[\mbox{bps/Hz}],\label{eq:sumRate}
\end{equation}
where $\alpha_{u}$ is a non-negative weight of the $u$th user.

\section{Precoder design\label{sec:Precoder-design}}

Limited CSI at the central coordination node makes the design of the
precoder all the more difficult. In this section, we derive the precoders
with limited CSI that also include the long term channel statistics
using the SINR definition from \eqref{eq:SINR_PL}.

\subsection{Successive second order cone programming}

We propose a SSOCP to solve the problem of precoder design with limited
information. SSOCP is based on SCA that allows us to efficiently solve
the problem with guaranteed convergence in every iteration. We adopt
an optimization framework originally proposed in \cite{VTT+14} for
linearizing a non-convex constraint that forms a constraint for the
useful signal. We also adopt the techniques in \cite{THT+12,HTT+13}
for handling the SINR, and reformulate as SOC constraints. The maximization
of weighted sum rate $R_{\mbox{tot}}$ with per-antenna power constraint%
\footnote{over all data symbols for a given channel realization%
} is formulated as
\begin{equation}
\begin{split}\underset{\mathbf{w}_{b,u}}{\mbox{maximize}} & \quad\underset{u}{\prod}\left(1+\gamma_{u}\right)^{\alpha_{u}}\\
\mbox{subject to} & \quad\underset{u\in\mathcal{U}_{b}}{\sum}|w_{b,u}^{(k)}|{}^{2}\leq P_{\mbox{max}},\forall b\in\mathcal{B}_{u},k=1,\ldots,N_{\text{T}},
\end{split}
\end{equation}
where the logarithm being a monotonically non-decreasing function
can be removed from the objective, and $P_{\mbox{max}}$ is the maximum
transmit power of an antenna of a BS serving a set of $\mathcal{U}_{b}$
users. This can be recast by letting $t_{u}=\left(1+\gamma_{u}\right)^{\alpha_{u}}$
where $\gamma_{u}$ is from \eqref{eq:SINR_PL} and adding a slack
variable $\beta_{u}$ as\begin{subequations}
\begin{eqnarray}
 & \underset{t_{u},\beta_{u},\mathbf{w}_{b,u}}{\mbox{maximize}} & \underset{u}{\prod}t_{u}\label{eq:MaxSR-1}\\
 & \mbox{subject to} & \frac{\left|\underset{b\in\mathcal{B}_{u}}{\sum}\mathbf{h}_{b,u}\mathbf{w}_{b,u}\right|^{2}}{\beta_{u}}\geq t_{u}^{1/\alpha_{u}}-1,\forall u\in\mathcal{U},\label{eq:sig_pwr}\\
 &  & \underset{i\neq u}{\sum}\left\{ \left|\underset{b\in\mathcal{B}_{i}\cap\mathcal{B}_{u}}{\sum}\mathbf{h}_{b,u}\mathbf{w}_{b,i}\right|^{2}\hspace{-0.2cm}+|\mathcal{B}_{i}\backslash\mathcal{B}_{u}|\underset{b\in\mathcal{B}_{i}\backslash\mathcal{B}_{u}}{\sum}\hspace{-0.3cm}\lambda_{b,u}^{2}||\mathbf{w}_{b,i}||_{2}^{2}\right\} +N_{0}\leq\beta_{u},\forall u\in\mathcal{U},\label{eq:IntCstr}\\
 &  & \underset{u\in\mathcal{U}_{b}}{\sum}|w_{b,u}^{(k)}|{}^{2}\leq P_{\mbox{max}},\forall b\in\mathcal{B}_{u},k=1,\ldots,N_{\text{T}}.\label{eq:PerAntennaPwrConstr}
\end{eqnarray}
\end{subequations} The LHS of \eqref{eq:sig_pwr} is of the form
quadratic over linear, which is convex function, and $t_{u}^{1/\alpha_{u}}$
is convex only when $0<\alpha_{u}\leq1$, and concave when $\alpha_{u}>1$.
Thus, the constraint is non-convex. A concave approximation of the
LHS can be obtained as in \cite[(6b)]{VTT+14}, so we define the following
expressions
\begin{eqnarray}
p_{u}\triangleq\Re\left\{ \underset{b\in\mathcal{B}_{u}}{\sum}\mathbf{h}_{b,u}\mathbf{w}_{b,u}\right\}  & \mbox{and} & q_{u}\triangleq\Im\left\{ \underset{b\in\mathcal{B}_{u}}{\sum}\mathbf{h}_{b,u}\mathbf{w}_{b,u}\right\} .\label{eq:TaylorP}
\end{eqnarray}
By applying the first order Taylor expansion for $\frac{\left(p_{u}^{2}+q_{u}^{2}\right)}{\beta_{u}}$
in LHS of \eqref{eq:sig_pwr} around the local point $\left\{ \widetilde{p}_{u},\widetilde{q}_{u},\widetilde{\beta}_{u}\right\} ,\forall u\in\mathcal{U},$
we get
\begin{eqnarray}
\frac{2\widetilde{p}_{u}}{\widetilde{\beta}_{u}}\left(p_{u}-\widetilde{p}_{u}\right)+\frac{2\widetilde{q}_{u}}{\widetilde{\beta}_{u}}\left(q_{u}-\widetilde{q}_{u}\right)+\frac{\widetilde{p}_{u}^{2}+\widetilde{q}_{u}^{2}}{\widetilde{\beta}_{u}}\left(1-\left(\frac{\beta_{u}-\widetilde{\beta}_{u}}{\widetilde{\beta}_{u}}\right)\right)+1 & \geq & t_{u}^{1/\alpha_{u}}.\label{eq:ReformTaylor}
\end{eqnarray}
When $\alpha_{u}>1$, $t_{u}^{1/\alpha_{u}}$ in the RHS of \eqref{eq:ReformTaylor}
is not convex, so it needs to be replaced by its upper bound. Doing
as in \cite{VTT+14}-\cite{HTT+13}, with the first order approximation
at the point $\widetilde{t}_{u}$, we obtain 
\begin{equation}
t_{u}^{1/\alpha_{u}}\leq\widetilde{t}_{u}^{1/\alpha_{u}}+\frac{1}{\alpha_{u}}\widetilde{t}_{u}^{\frac{1}{\alpha_{u}}-1}\left(t_{u}-\widetilde{t}_{u}\right).
\end{equation}
Therefore, combining with \eqref{eq:ReformTaylor}, we get
\begin{eqnarray}
\frac{2\widetilde{p}_{u}}{\widetilde{\beta}_{u}}\left(p_{u}-\widetilde{p}_{u}\right)+\frac{2\widetilde{q}_{u}}{\widetilde{\beta}_{u}}\left(q_{u}-\widetilde{q}_{u}\right)+\frac{\widetilde{p}_{u}^{2}+\widetilde{q}_{u}^{2}}{\widetilde{\beta}_{u}}\left(1-\left(\frac{\beta_{u}-\widetilde{\beta}_{u}}{\widetilde{\beta}_{u}}\right)\right)+1\nonumber \\
\geq\widetilde{t}_{u}^{1/\alpha_{u}}+\frac{1}{\alpha_{u}}\widetilde{t}_{u}^{\frac{1}{\alpha_{u}}-1}\left(t_{u}-\widetilde{t}_{u}\right).\label{eq:ReformTaylorFinal}
\end{eqnarray}
Now consider \eqref{eq:IntCstr} which can be rewritten as an SOC
constraint \cite{HTT+13}
\begin{eqnarray}
\left(\underset{i\neq u}{\sum}\left\{ \left|\underset{b\in\mathcal{B}_{i}\cap\mathcal{B}_{u}}{\sum}\mathbf{h}_{b,u}\mathbf{w}_{b,i}\right|^{2}+|\mathcal{B}_{i}\backslash\mathcal{B}_{u}|\underset{b\in\mathcal{B}_{i}\backslash\mathcal{B}_{u}}{\sum}\lambda_{b,u}^{2}||\mathbf{w}_{b,i}||_{2}^{2}\right\} +\left(\sqrt{N_{0}}\right)^{2}+\frac{1}{4}\left(\beta_{u}-1\right)^{2}\right)^{1/2}\nonumber \\
\leq\frac{1}{2}\left(\beta_{u}+1\right),\forall u\in\mathcal{U}.\label{eq:SOC_IntCtr-1}
\end{eqnarray}
Therefore, the reformulated convex problem for precoder design with
the objective of maximizing the geometric mean of $t_{u}$ becomes\raggedbottom
\begin{equation}
\begin{split}\underset{t_{u},\beta_{u},\mathbf{w}_{b,u}}{\mbox{maximize}} & \quad\left(\overset{\left|\mathcal{U}\right|}{\underset{u=1}{\prod}}t_{u}\right)^{1/\left|\mathcal{U}\right|}\\
\mbox{subject to} & \hspace{0.55cm}\eqref{eq:PerAntennaPwrConstr},\eqref{eq:ReformTaylorFinal}\,\,\mbox{and}\,\,\eqref{eq:SOC_IntCtr-1},
\end{split}
\label{eq:GaneshCVX}
\end{equation}
where the geometric mean is concave, and the exponent does not affect
the optimal value. This is performed merely to simplify the implementation.
Also, the interfering terms can be collected in a vector as
\begin{equation}
\mathbf{r}_{i}=\left[\begin{array}{c}
\underset{b\in\mathcal{B}_{i}\cap\mathcal{B}_{u}}{\sum}\mathbf{h}_{b,u}\mathbf{w}_{b,i}\\
\sqrt{|\mathcal{B}_{i}\backslash\mathcal{B}_{u}|}\lambda_{b',u}\mathbf{w}_{b',i}
\end{array}\right],b'\in\mathcal{B}_{i}\backslash\mathcal{B}_{u},\forall i\neq u.
\end{equation}
The SSOCP with the above simplified notation is summarized in Algorithm~\ref{alg:SSOCP-algorithm-for}.
\begin{algorithm}[H]
\begin{centering}
\protect\caption{\label{alg:SSOCP-algorithm-for}SSOCP algorithm for precoder design}

\par\end{centering}

\small{

\begin{algorithmic}[1]

\STATE To avoid numerical instability, rescale the aggregated channel
matrix and the noise power with a factor of the least pathloss such
that the SINR is the same. 

\STATE Set $maxRetries=\mbox{MAXRETRIES}$, see Fig.~\ref{fig:A-comparison-of}
for a possible choice.

\WHILE {$maxRetries$ }

\STATE \label{-Random-initialization}Randomly initialize the non-zero
precoding weight, $\mathbf{w}_{b,u}$, from $\mathcal{CN}(0,1)$,
and ensure the power of each antenna is limited to $P_{\mbox{max}}$.

\STATE Calculate $\gamma_{u}$ based on \eqref{eq:SINR_PL}, $\forall u$.

\STATE Set $n=0$ 

\STATE Evaluate $\widetilde{p}_{u}^{(n)}$ and $\widetilde{q}_{u}^{(n)}$
from \eqref{eq:TaylorP}.

\STATE Evaluate $t_{u}^{(n)}=\left(1+\gamma_{u}\right)^{\alpha_{u}}$
and $\beta_{u}^{(n)}=\frac{\left(\widetilde{p}_{u}^{(n)}\right)^{2}+\left(\widetilde{q}_{u}^{(n)}\right)^{2}}{t_{u}^{(n)}-1}$

\STATE Set $maxIter=\mbox{MAXITER}$

\WHILE {$maxIter$ AND $\dagger$}

\STATE Treat $p_{u}^{(n)}$ and $q_{u}^{(n)}$ as \emph{expressions
}in CVX \cite{GB11} which will be used in \eqref{eq:ReformTaylorFinal}.

\STATE \label{-Solve-theGaneshSCAProblem_START}Solve the convex
problem \eqref{eq:GaneshCVX} as
\[
\begin{split}\underset{t_{u},\beta_{u},\mathbf{w}_{b,u}}{\mbox{\textbf{maximize}}} & \quad\mbox{\textbf{geo\_mean}}\left(t_{u}\right)\\
\mbox{subject to}\\
 & \quad\left|\left|\begin{array}{c}
\mathbf{r}_{i}\\
\sqrt{N_{0}}\\
\frac{1}{2}\left(\beta_{u}-1\right)
\end{array}\right|\right|_{2}\leq\frac{1}{2}\left(\beta_{u}+1\right),\\
 & \quad\forall i\in\mathcal{U},\\
 & \quad\eqref{eq:PerAntennaPwrConstr},\\
 & \quad\mbox{and}\,\eqref{eq:ReformTaylorFinal},\forall u\in\mathcal{U}.
\end{split}
\]

\STATE Update: $t_{u}^{(n+1)}=t_{u}^{(n)},\beta_{u}^{(n+1)}=\beta_{u}^{(n)}$ 

\STATE Update: $p_{u}^{(n+1)}=p_{u}^{(n)},q_{u}^{(n+1)}=q_{u}^{(n)}$

\STATE Update: $n=n+1$

\STATE $maxIter=maxIter-1$

\STATE Evaluate and save the best weighted sum rate achieved so far,
as well as the corresponding precoding weights.

\ENDWHILE

\STATE $maxRetries=maxRetries-1$

\ENDWHILE

\end{algorithmic}

}

$\dagger$ The weighted sum rate does not improve within a certain
tolerance.
\end{algorithm}
The weighted sum rate maximization is a non-convex problem, and the
solution may end up as an inefficient local optimum. In order to further
improve the solution, we introduce random initialization similar to
\cite{TCJ08}, where we select the best solution out of a number of
random initialization. For a given aggregated channel matrix, a small
increase in the number of random initializations, as in step~\ref{-Random-initialization},
increases the probability to find a solution close to the global optimal
\cite{TCJ08}. 

The convergence of the proposed SSOCP algorithm closely follows the
analysis carried out for the full CSI case in \cite{VTT+14}-\cite{HTT+13}.
Reformulating the SINR constraints to cope with incomplete CSI does
not affect the convergence of the SCA. The interference and noise
terms are transformed into a SOC from \eqref{eq:IntCstr}, and the
convex function in \eqref{eq:sig_pwr} is approximated with a linear
lower bound at each iteration. This results in an SCA for every iteration,
where the objective is monotonically non-decreasing, thereby guaranteeing
convergence. In the subsequent section, we apply the branch and bound
technique to show that the proposed SSOCP is very close to the optimal.

\subsection{Optimization via weighted mean square error minimization}

Maximizing the weighted sum rate can be equivalently formulated as
minimizing the weighted sum MSE \cite{SR+11,CAC+08}. In this section,
we extend this result to the case of limited CSI and efficient backhauling.
We derive the precoder based on weighted MSE formulation that accounts
for using long term channel statistics in the precoder design when
there is limited CSI at the central coordination node. Consider the
received signal at the $u$th user as in \eqref{eq:received sig}.
The estimated $u$th user data at the receiver is $\hat{x}_{u}=a_{u}y_{u}$,
where $a_{u}\in\mathbb{C}$ is the receiver weight. The MSE at the
$u$th receiver $\xi_{u}$, can be formulated as
\begin{eqnarray}
\xi_{u} & = & E_{x_{u},n_{u}}\left[\left(x_{u}-\hat{x}_{u}\right)\left(x_{u}-\hat{x}_{u}\right)^{H}\right]\nonumber \\
 & = & E_{x_{u},n_{u}}\left[\left(x_{u}-a_{u}y_{u}\right)\left(x_{u}-a_{u}y_{u}\right)^{H}\right]\nonumber \\
 & = & E_{x_{u}}\left[x_{u}x_{u}^{H}\right]-a_{u}^{H}E_{x_{u},n_{u}}\left[x_{u}y_{u}^{H}\right]-a_{u}E_{x_{u},n_{u}}\left[y_{u}x_{u}^{H}\right]+a_{u}a_{u}^{H}E_{x_{u},n_{u}}\left[y_{u}y_{u}^{H}\right]\nonumber \\
 & = & 1-a_{u}^{H}\underset{b\in\mathcal{B}_{u}}{\sum}\left(\mathbf{h}_{b,u}\mathbf{w}_{b,u}\right)^{H}-a_{u}\underset{b\in\mathcal{B}_{u}}{\sum}\mathbf{h}_{b,u}\mathbf{w}_{b,u}+a_{u}a_{u}^{H}\widetilde{c}_{u}.\label{eq:MSE_before_rxFilter}
\end{eqnarray}
where $E_{x_{u}}\left[x_{u}x_{u}^{H}\right]=1$ as the user data is
zero mean with unit power and $\widetilde{c}_{u}=E_{x_{u},n_{u}}\left[y_{u}y_{u}^{H}\right]$
is the variance of the received signal which can be evaluated as 
\begin{eqnarray}
\widetilde{c}_{u} & = & E_{x_{u},n_{u}}\left[y_{u}y_{u}^{H}\right]\nonumber \\
 & = & \underset{b\in\mathcal{B}_{u}}{\sum}\mathbf{h}_{b,u}\mathbf{w}_{b,u}\left(\mathbf{h}_{b,u}\mathbf{w}_{b,u}\right)^{H}+\underset{i\neq u}{\sum}\underset{b\in\mathcal{B}_{i}}{\sum}\mathbf{h}_{b,u}\mathbf{w}_{b,i}\left(\mathbf{h}_{b,u}\mathbf{w}_{b,i}\right)^{H}+E_{n_{u}}\left[n_{u}n_{u}^{H}\right]\nonumber \\
 & = & \underset{\forall i}{\sum}\underset{b\in\mathcal{B}_{u}}{\sum}\mathbf{h}_{b,u}\mathbf{w}_{b,i}\left(\mathbf{h}_{b,u}\mathbf{w}_{b,i}\right)^{H}+N_{0},\label{eq:RxCov}
\end{eqnarray}
where $E_{n_{u}}\left[n_{u}n_{u}^{H}\right]=N_{0}$. Similar to \eqref{eq:SINR_PL},
we split the interference terms to incorporate the long term channel
statistics for the unknown channel components, and bound the variance
to be pessimistic as
\begin{eqnarray}
c_{u} & = & N_{0}+\left|\underset{b\in\mathcal{B}_{u}}{\sum}\mathbf{h}_{b,u}\mathbf{w}_{b,u}\right|^{2}+\underset{i\neq u}{\sum}\left\{ \left|\underset{b\in\mathcal{B}_{i}\cap\mathcal{B}_{u}}{\sum}\mathbf{h}_{b,u}\mathbf{w}_{b,i}\right|^{2}+|\mathcal{B}_{i}\backslash\mathcal{B}_{u}|\underset{b\in\mathcal{B}_{i}\backslash\mathcal{B}_{u}}{\sum}\lambda_{b,u}^{2}||\mathbf{w}_{b,i}||_{2}^{2}\right\} .\label{eq:RxCovWithLambda}
\end{eqnarray}
The detailed steps are in Appendix~\ref{AppRxCov}. Therefore, the
MSE in \eqref{eq:MSE_before_rxFilter} has $\widetilde{c}_{u}$ replaced
with $c_{u}$. To find the optimal receive weight, $a_{u}^{\star}=a_{u}^{\text{MMSE}}$,
we need to take the $\bigtriangledown_{a_{u}}\left(\xi_{u}\right)=0$,
which works out to be
\begin{equation}
a_{u}^{\star}=\underset{b\in\mathcal{B}_{u}}{\sum}\left(\mathbf{h}_{b,u}\mathbf{w}_{b,u}\right)^{H}c_{u}^{-1}.\label{eq:RxFilter}
\end{equation}
Therefore, the MMSE is
\begin{eqnarray}
\bar{\xi_{u}} & = & 1-\underset{b\in\mathcal{B}_{u}}{\sum}\left(\mathbf{h}_{b,u}\mathbf{w}_{b,u}\right)^{H}c_{u}^{-1}\underset{b\in\mathcal{B}_{u}}{\sum}\mathbf{h}_{b,u}\mathbf{w}_{b,u}\nonumber \\
 & = & 1-a_{u}^{\star}\underset{b\in\mathcal{B}_{u}}{\sum}\mathbf{h}_{b,u}\mathbf{w}_{b,u}=\frac{1}{1+\gamma_{u}}.\label{eq:MMSEOpt}
\end{eqnarray}
Thus, maximizing the weighted sum rate \eqref{eq:sumRate} can be
formulated equivalent to a $\mbox{log}(\mbox{MSE})$ minimization
problem \cite{SR+11,CAC+08} as
\begin{equation}
\begin{split}\underset{\mathbf{w}_{b,u}}{\mbox{minimize}} & \quad\underset{u}{\sum}\alpha_{u}\mbox{log}_{2}\bar{\xi_{u}}\\
\mbox{subject to} & \quad\eqref{eq:PerAntennaPwrConstr}.
\end{split}
\label{eq:MSE_NonCVX}
\end{equation}
 where $\alpha_{u}$ is a non-negative weight. The problem \eqref{eq:MSE_NonCVX}
is non-convex, so as a first step to find a tractable local solution
we introduce a new variable, $\breve{\xi}_{u}$, to upper bound the
MSE as well as treat the receive scalar, $a_{u}$, as an optimization
variable. The reformulated problem is \begin{subequations}
\begin{eqnarray}
 & \underset{\mathbf{w}_{b,u},a_{u},\breve{\xi}_{u}}{\mbox{minimize}} & \underset{u}{\sum}\alpha_{u}\mbox{log}_{2}\breve{\xi}_{u}\label{eq:MSE_UpperBound}\\
 & \mbox{subject to} & 1-a_{u}^{H}\underset{b\in\mathcal{B}_{u}}{\sum}\left(\mathbf{h}_{b,u}\mathbf{w}_{b,u}\right)^{H}-a_{u}\underset{b\in\mathcal{B}_{u}}{\sum}\mathbf{h}_{b,u}\mathbf{w}_{b,u}+a_{u}a_{u}^{H}c_{u}\leq\breve{\xi}_{u},\forall u,\label{eq:MSE_upperBoundConstraint}\\
 &  & \eqref{eq:PerAntennaPwrConstr}.
\end{eqnarray}
\end{subequations}Still the MSE constraint \eqref{eq:MSE_upperBoundConstraint}
is not jointly convex with respect to both $\mathbf{w}_{b,u}$ and
$a_{u}$. However, for a fixed receiver, $a_{u}$, the constraint
becomes convex. A successive linear approximation of the concave objective
is carried out at the point $\widetilde{\xi}_{u}^{(n)}$ in the $n$th
iteration as 
\begin{equation}
\mbox{log}_{2}\breve{\xi}_{u}^{(n)}\approx\mbox{log}_{2}\widetilde{\xi}_{u}^{(n)}+d_{u}^{(n)}\left(\breve{\xi}_{u}^{(n)}-\widetilde{\xi}_{u}^{(n)}\right)\mbox{log}_{2}e,\label{eq:LinearizeMSE}
\end{equation}
where $d_{u}^{(n)}$ is the linearizing coefficient, which is a non-negative
MSE weight for the $n$th iteration, and it works out to be
\begin{equation}
d_{u}^{(n)}=\frac{1}{\widetilde{\xi}_{u}^{(n)}}.\label{eq:RxWeight}
\end{equation}
Subsituting \eqref{eq:LinearizeMSE} in the objective \eqref{eq:MSE_UpperBound},
and considering only those terms that depend on $\mathbf{w}_{b,u}$
for a fixed $a_{u}$, while ignoring the constant terms and iteration
index in \eqref{eq:LinearizeMSE}, the objective becomes
\begin{equation}
\underset{\mathbf{w}_{b,u},\breve{\xi}_{u}}{\mbox{minimize}}\quad\underset{u}{\sum}\alpha_{u}d_{u}\breve{\xi}_{u}.\label{eq:EvilXi}
\end{equation}
Furthermore, the MSE constraint is tight at the optimal solution.
Thus, we replace the MSE upper bound with the actual MSE expression
as 
\begin{eqnarray}
\zeta & = & \underset{u}{\sum}\alpha_{u}d_{u}\xi_{u}\label{eq:LinPetriObj}\\
 & = & \underset{u}{\sum}-2\alpha_{u}\Re\left\{ d_{u}a_{u}\underset{b\in\mathcal{B}_{u}}{\sum}\mathbf{h}_{b,u}\mathbf{w}_{b,u}\right\} +\underset{u}{\sum}\alpha_{u}d_{u}a_{u}\left\{ \left|\underset{b\in\mathcal{B}_{u}}{\sum}\mathbf{h}_{b,u}\mathbf{w}_{b,u}\right|^{2}\right.\nonumber \\
 &  & \left.+\underset{i\neq u}{\sum}\left\{ \left|\underset{b\in\mathcal{B}_{i}\cap\mathcal{B}_{u}}{\sum}\mathbf{h}_{b,u}\mathbf{w}_{b,i}\right|^{2}+|\mathcal{B}_{i}\backslash\mathcal{B}_{u}|\underset{{\scriptstyle b\in\mathcal{B}_{i}\backslash\mathcal{B}_{u}}}{\sum}\lambda_{b,u}^{2}||\mathbf{w}_{b,i}||_{2}^{2}\right\} \right\} .\label{eq:ObjMMSE}
\end{eqnarray}
For fixed $a_{u}$, (21) can be solved via successive linearization
until convergence. For fixed $\mathbf{w}_{b,u}$, the optimal solution
of (21) is given by the optimal receiver weight, $a_{u}$ in \eqref{eq:RxFilter}.
This leads to alternating optimization with monotonic convergence.
In practice, we update $a_{u}$ and then $d_{u}$ just once without
sacrificing monotonicity of the objective. Thus, the original problem
(21) can be split as a 3-stage algorithm \cite{SR+11,CAC+08,KA+13},
where the receiver weights, linearizing coefficients, and the precoders
are optimized in an alternating manner, i.e., (i) the receiver weights
are updated for a given precoder, (ii) the linearizing coefficients
are updated for a given precoder, and (iii) the precoders are evaluated
for the given receiver weights and linearizing coefficients. When
compared to \cite{SR+11,CAC+08,KA+13}, we design the precoder with
limited information and achieve efficient backhauling in a JT-CoMP
scenario with per-antenna power constraint. The MSE based precoder
design with limited information is outlined in Algorithm~\ref{alg:The-MSE-approach}.
The convergence is evaluated based on the MSE of each user as $-\underset{u}{\sum}\alpha_{u}\mbox{log\ensuremath{{}_{2}}}\xi_{u}$.
This is a monotonically non-decreasing function, and the algorithm
is terminated when there is no further improvement. 
\begin{algorithm}[H]
\begin{centering}
\protect\caption{\label{alg:The-MSE-approach}The MSE approach for finding the precoder}

\par\end{centering}

\small{

\begin{algorithmic}[1]

\STATE Randomly initialize every precoding weight, $\mathbf{w}_{b,u}$,
from $\mathcal{CN}(0,1)$, apply the equivalent limited backhauling,
and ensure the power of each antenna is limited to $P_{\mbox{max}}$.

\WHILE { Convergence } \label{-Convergence-MSEConvergence}

\STATE Evaluate: $a_{u},d_{u},c_{u},\forall u$

\STATE Solve the convex problem as
\[
\begin{split}\underset{t_{u},\mathbf{w}_{b,u}}{\mbox{\textbf{minimize}}}\quad & \underset{u}{\sum}t_{u}\\
\mbox{subject to}\quad & -2\alpha_{u}\Re\left\{ d_{u}a_{u}\underset{b\in\mathcal{B}_{u}}{\sum}\mathbf{h}_{b,u}\mathbf{w}_{b,u}\right\} \\
 & +\alpha_{u}d_{u}a_{u}a_{u}^{H}\left|\underset{b\in\mathcal{B}_{u}}{\sum}\mathbf{h}_{b,u}\mathbf{w}_{b,u}\right|^{2}\\
 & +\alpha_{u}d_{u}a_{u}a_{u}^{H}\left\{ \underset{i\neq u}{\sum}\left\{ \left|\underset{b\in\mathcal{B}_{i}\cap\mathcal{B}_{u}}{\sum}\mathbf{h}_{b,u}\mathbf{w}_{b,i}\right|^{2}\right.\right.\\
 & \left.\left.+|\mathcal{B}_{i}\backslash\mathcal{B}_{u}|\underset{b\in\mathcal{B}_{i}\backslash\mathcal{B}_{u}}{\sum}\lambda_{b,u}^{2}||\mathbf{w}_{b,i}||_{2}^{2}\right\} \right\} \leq t_{u},\forall u\in\mathcal{U},\\
 & \eqref{eq:PerAntennaPwrConstr}.
\end{split}
\]

Based on precoding weights, steps \ref{-Update:-UpdateMSEStart}-\ref{-Update:-UpdateMSEEnd}
below are applied $\forall u$.

\STATE \label{-Update:-UpdateMSEStart}Update: receiver variance,
$c_{u}$, based on \eqref{eq:RxCovWithLambda}

\STATE Update: receiver weight, $a_{u}$, based on \eqref{eq:RxFilter}

\STATE \label{-Update:-UpdateMSEEnd}Update: linearizing coefficient,
$d_{u}$, based on \eqref{eq:RxWeight}

\ENDWHILE 

\RETURN { Precoding matrix}

\end{algorithmic}

}
\end{algorithm}

The problem of minimizing, $\zeta$, can be solved either with generic
solvers, such as those provided with the CVX package \cite{GB11},
where the per-antenna power constraint is formulated as an SOC program,
or solving iteratively via the Karush-Kuhn-Tucker conditions as highlighted
in \cite[Appx. A]{KA+13}. Note that the latter approach may be preferrable
when the number of power constraints (corresponding to the dual variables)
is relatively small. The MSE algorithm guarantees convergence as shown
in \cite[Thm. 3]{SR+11}. The main difference to \cite{SR+11} is
that here the receive variance is affected by the long term channel
statistics, thus the same convergence analysis applies.

\subsection{Stochastic optimization using particle swarm optimization}

The researchers modeling the movement of birds or a shoal of fish
discovered that these movements were indeed performing optimization.
This gave birth to an entire field of swarm intelligence. In particular,
we focus on PSO, a stochastic optimization technique that can provide
an acceptable solution even when the problem is non-convex. PSO was
proposed to design the precoder in a CoMP setup with limited information
\cite{LBS12}. We consider the implementation of PSO as described
in \cite[Algo. 2]{LBS12}, with the addition of random initialization,
giving rise to a multi-start PSO such that global optimization can
be performed. PSO is a very attractive tool for precoder design as
it does not involve any matrix inversion, and the overhead of book
keeping the number of variables is very little compared to genetic
algorithms. However, being heuristic in nature, it does not guarantee
optimality and the algorithm might not converge in polynomial time
with the increase in problem size.

\subsection{Branch and Bound\label{sub:Branch-and-Bound}}

The SSOCP and the MSE reformulated approaches are iterative algorithms
where every sub-step is optimal and well justified, leading to monotonic
improvement of the objective, with guaranteed convergence. However,
these approaches even with a large number of random initializations
is not guaranteed to obtain the optimal solution, as the problem is
non-convex and NP-hard. Hence, we need to verify that our proposed
solution is tightly bounded. In this regard, we consider \cite{JWC+12}
where branch and bound (BB) is applied for weighted sum rate maximization
in MISO downlink cellular networks. We reformulate this problem for
joint transmission CoMP. This provides the lower and upper bounds
for the problem with full and limited feedback, and also when the
pessimistic statistical interference model is used in the precoder
design. With this approach, we can say how close the proposed algorithm
is from being globally optimum. 

When limited CSI information is available at the central coordination
node, the branch and bound technique can be applied via reformulating
the weighted sum rate maximization in MISO downlink \cite{JWC+12}
to joint transmission CoMP networks. It is intuitive to observe that
the reformulation is exactly the same as \cite{JWC+12} however the
SINRs are based on \eqref{eq:SINR_PL}. Instead of rewriting the whole
branch and bound procedure as described in \cite{JWC+12}, we highlight
the main differences in the reformulated problem. The initialization
of the hyperrectangle in \cite[(9)]{JWC+12} is 
\begin{equation}
\mathcal{Q}_{\text{init}}=\left\{ \mathbf{\gamma}|0\leq\gamma_{u}\leq\left|\mathcal{B}_{u}\right|N_{\text{T}}P_{\max}\underset{b\in\mathcal{B}_{u}}{\sum}||\mathbf{h}_{b,u}||_{2}^{2}/N_{0}{\scriptstyle ,\forall u\in\mathcal{U}}\right\} .\label{eq:Qinit}
\end{equation}
The initial hyperrectangle, $Q\in\mathcal{Q}_{\text{init}}$ comprises
of $\gamma_{\text{max}}\in\mathbb{R}_{0}^{+}$, and $\gamma_{\text{min}}\in\mathbb{R}_{0}^{+},\gamma_{\text{min}}=\mathbf{0}_{\left|\mathcal{U}\right|}$.
The upper limit in \eqref{eq:Qinit} can be obtained for each user,
when considering only the SNR from \eqref{eq:SINR_PL}, where the
$u$th user is the only user in the system without any interference.
It is intuitive to see that using the Cauchy-Schwartz inequality we
have $\left|\underset{b\in\mathcal{B}_{u}}{\sum}\mathbf{h}_{b,u}\mathbf{w}_{b,u}\right|^{2}\leq\left|\mathcal{B}_{u}\right|\underset{b\in\mathcal{B}_{u}}{\sum}\left|\mathbf{h}_{b,u}\mathbf{w}_{b,u}\right|^{2}\le\left|\mathcal{B}_{u}\right|\underset{b\in\mathcal{B}_{u}}{\sum}||\mathbf{h}_{b,u}||_{2}^{2}||\mathbf{w}_{b,u}||_{2}^{2}\leq\left|\mathcal{B}_{u}\right|N_{\text{T}}P_{\max}\underset{b\in\mathcal{B}_{u}}{\sum}||\mathbf{h}_{b,u}||_{2}^{2}$.
The last inequality is due to the per-antenna power constraint. The
other main contribution lies in the check for the feasibility under
limited feedback and backhauling constraint. This is captured in Algorithm~\ref{alg:-Check-if}.
The feasibility check is performed as part of the BB technique, and
for completeness we provide a cookbook version of BB in Appendix~\ref{AppCookboockBB}.

\begin{algorithm}[H]
\begin{centering}
\protect\caption{\label{alg:-Check-if} Check if $\gamma=\left[\gamma_{1},\ldots,\gamma_{u},\ldots,\gamma_{\left|\mathcal{U}\right|}\right]$
is feasible.}

\par\end{centering}

\small{

\begin{algorithmic}[1]

\STATE Check for feasibility by solving the convex problem
\[
\begin{split}\mbox{\textbf{find}} & \quad\mathbf{w}_{b,u},\forall b\in\underset{k\in\mathcal{U}}{\cup}\mathcal{B}_{k},\forall u\in\mathcal{U},\\
\mbox{subject to}\\
 & \quad\left|\left|\begin{array}{c}
\underset{b\in\mathcal{B}_{u}}{\sum}\mathbf{h}_{b,u}\mathbf{w}_{b,u}\\
\mathbf{r}_{i}\\
\sqrt{N_{0}}
\end{array}\right|\right|_{2}\leq\sqrt{1+\frac{1}{\gamma_{u}}}\underset{b\in\mathcal{B}_{u}}{\sum}\mathbf{h}_{b,u}\mathbf{w}_{b,u},\\
 & \quad\forall i\in\mathcal{U},\\
 & \quad\eqref{eq:PerAntennaPwrConstr}.
\end{split}
\ddagger
\]

\RETURN { feasibility, and save $\mathbf{w}_{b,u},\forall b,u$ when
feasible}

\end{algorithmic}

}

$\ddagger$ The RHS of this SOCP formulation can be argued along the
same lines as~\cite[Sec. IV.B]{WES06}.
\end{algorithm}

\section{Simulations\label{sec:Simulations}}

We consider $\left|\mathcal{U}\right|=3$ users that are uniformly
dropped around the cell-edge at the intersection of $\left|\mathcal{B}\right|=3$
BSs, where each BS has $N_{\text{T}}=1,3$ transmit antenna(s) covering
a cell-radius of 500 m. The cell-edge SNR is defined as the SNR experienced
by one user at the cell-edge. For simplicity, we set $\alpha_{u}=1,\forall u$.
The variance of the shadow fading component is $\sigma_{\text{SF}}^{2}=8\,\mbox{dB}$.
The receiver noise power is $N_{0}=kTB_{n}$ Watts, where $k$ is
the Boltzmann's constant $1.38\times10^{-23}$Joules/Kelvin, $T=290$
Kelvin is the operating temperature, and $B_{n}=10\,\mbox{MHz}$ is
the system bandwidth.

The legends in the following figures are summarized in Table~\ref{tab:The-interpretation-of}.
They capture as to how much feedback or backhauling is required or
being used based on a given relative threshold. The algorithms without
subscripts such as SSOCP, PSO, MSE, ZF, $\mbox{BB}_{\text{UB}}$ and
$\mbox{BB}_{\text{LB}}$, capture the case of full feedback and full
backhauling, when the relative threshold, $T=\infty\,\mbox{dB}$.
The algorithms with subscripts such as $\mbox{SSOCP}_{\lambda,\text{PL,0}}$,
$\mbox{MSE}_{\lambda,\text{PL,0}}$, $\mbox{BB}_{\lambda,\text{UB,0}}$
and $\mbox{BB}_{\lambda,\text{LB,0}}$ capture the case of limited
feedback incorporating our proposed long term channel statistics,
and with limited backhauling. The $\mbox{SSOCP}_{\text{PL,0}}$ algorithm
captures the naive approach of including the long term channel statistics
where they are directly replacing the missing channel coefficients
in the interference terms of SINR formulation, when there is limited
feedback and limited backhauling. The algorithms with subscripts such
as $\mbox{SSOCP}_{\text{0}}$, $\mbox{PSO}_{\text{0}}$, $\mbox{BB}_{\text{UB,0}}$
and $\mbox{BB}_{\text{LB,0}}$ capture the case of limited feedback
and limited backhauling without the use of any side information.

\begin{table}[tbh]
\protect\caption{\label{tab:The-interpretation-of}The interpretation of legends related
to the information available and generated at the central coordination
node is listed below.}

\begin{centering}
\begin{tabular}{|c|c|c|}
\hline 
$\mbox{Legend}^{\dagger}$ & CSI Feedback & Precoding weights\tabularnewline
\hline 
\hline 
$\mbox{BB}_{\text{LB}}$ & Full & Full\tabularnewline
\hline 
$\mbox{BB}_{\text{UB}}$ & Full & Full\tabularnewline
\hline 
$\mbox{BB}_{\lambda,\text{LB,0}}$ & Use long term stats \eqref{eq:SINR_PL} & Limited\tabularnewline
\hline 
$\mbox{BB}_{\lambda,\text{UB,0}}$ & Use long term stats \eqref{eq:SINR_PL} & Limited\tabularnewline
\hline 
$\mbox{BB}_{\text{LB,0}}$ & Limited & Limited\tabularnewline
\hline 
$\mbox{BB}_{\text{UB,0}}$ & Limited & Limited\tabularnewline
\hline 
MSE & Full & Full\tabularnewline
\hline 
$\mbox{MSE}_{\lambda,\text{PL,0}}$ & Use long term stats \eqref{eq:ObjMMSE} & Limited\tabularnewline
\hline 
PSO & Full & Full\tabularnewline
\hline 
$\mbox{PSO}_{\text{0}}$ & Limited & Limited\tabularnewline
\hline 
SSOCP & Full & Full\tabularnewline
\hline 
$\mbox{SSOCP}_{\lambda,\text{PL,0}}$ & Use long term stats \eqref{eq:SINR_PL} & Limited\tabularnewline
\hline 
$\mbox{SSOCP}_{\text{PL,0}}$ & Use long term stats directly & Limited\tabularnewline
\hline 
$\mbox{SSOCP}_{\text{0}}$ & Limited & Limited\tabularnewline
\hline 
ZF & Full & Full\tabularnewline
\hline 
\end{tabular}\\
~
\par\end{centering}

$\dagger$The acronyms in the legend are summarized here for convenience.
The $\mbox{BB}_{\text{UB}}$ and $\mbox{BB}_{\text{LB}}$ denote the
upper and lower bound obtained from branch and bound as presented
in Algorithm~\ref{alg:The-branch-and}, PSO: particle swarm optimization,
MSE: weighted mean square error, SSOCP: successive second order cone
programming, and ZF: zero forcing.
\end{table}

\subsection{Effect of threshold and cell-edge SNR}

\begin{figure}[tbh]
\begin{centering}
\includegraphics[width=0.4\paperwidth]{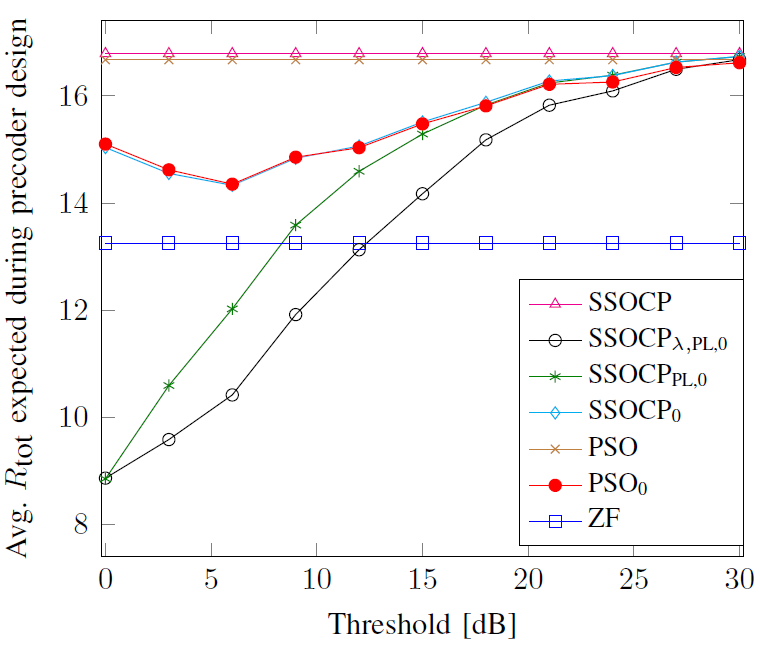}
\par\end{centering}

\protect\caption{\label{fig:With-PL+SF-1} The performance of the precoders in terms
of average weighted sum rate, $R_{\text{tot}}$ expected when designing
the precoder at the central coordination node. The cell-edge SNR is
15~dB, $N_{\text{T}}=1$, $\left|\mathcal{B}\right|=3$, and $\left|\mathcal{U}\right|=3$. }
\end{figure}
\begin{figure}[tbh]
\begin{centering}
\includegraphics[width=0.4\paperwidth]{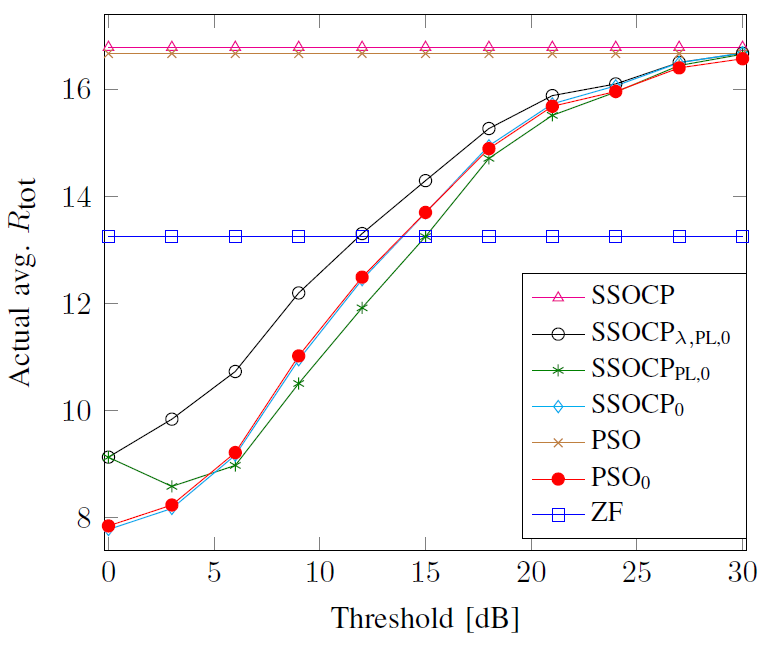}
\par\end{centering}

\protect\caption{\label{fig:With-PL+SF} The performance of the precoders in terms
of the actual average weighted sum rate, $R_{\text{tot}}$ evaluated
due to the transmission to the users, with increasing threshold for
a given cell-edge SNR of 15~dB, $N_{\text{T}}=1$, $\left|\mathcal{B}\right|=3$,
and $\left|\mathcal{U}\right|=3$. }
\end{figure}
Fig.~\ref{fig:With-PL+SF-1} shows the \emph{expected} average weighted
sum rate evaluated at the central coordination node when designing
the precoder for various relative thresholds. It is intuitive to note
that with complete information the performance of SSOCP, PSO and ZF
are independent of the threshold. With limited information, $\mbox{PSO}_{0}$
and $\mbox{SSOCP}_{0}$ have similar performance. The most interesting
curves are $\mbox{SSOCP}_{\lambda,\text{PL},0}$ and $\mbox{SSOCP}_{\text{PL},0}$,
where $\mbox{SSOCP}_{\lambda,\text{PL},0}$ incorporates the long
term channel statistics in the interference terms, thereby resulting
in a pessimistic precoder design. The $\mbox{SSOCP}_{\text{PL},0}$
naively replaces zeros with these long term channel statistics and
appear to achieve superior performance when designing the precoder
at the central coordination node. However, it is important to observe
the actual performance of the precoder due to the transmission to
the user. This is captured in Fig.~\ref{fig:With-PL+SF} where $\mbox{SSOCP}_{\lambda,\text{PL},0}$
outperforms all other cases when there is limited information. It
is interesting to note that the expected rates in Fig.~\ref{fig:With-PL+SF-1}
are in line with the actual rates in Fig.~\ref{fig:With-PL+SF} for
the proposed $\mbox{SSOCP}_{\lambda,\text{PL},0}$ approach.

\begin{figure}[tbh]
\begin{centering}
\includegraphics[width=0.4\paperwidth]{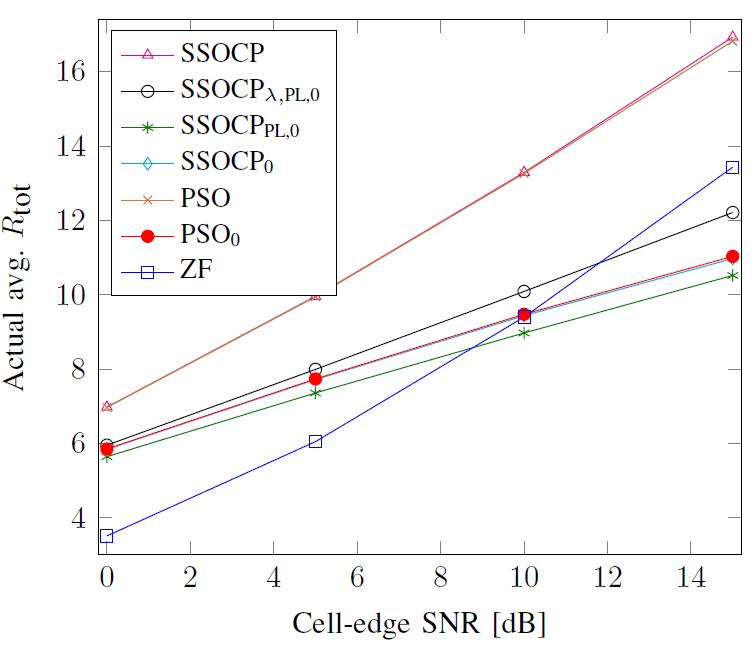}
\par\end{centering}

\protect\caption{\label{fig:Average-Sum-Rate-1}Average weighted sum rate versus cell-edge
SNRs for a relative thresholds of 9~dB.}
\end{figure}
\begin{figure}[tbh]
\centering{}\includegraphics[width=0.4\paperwidth]{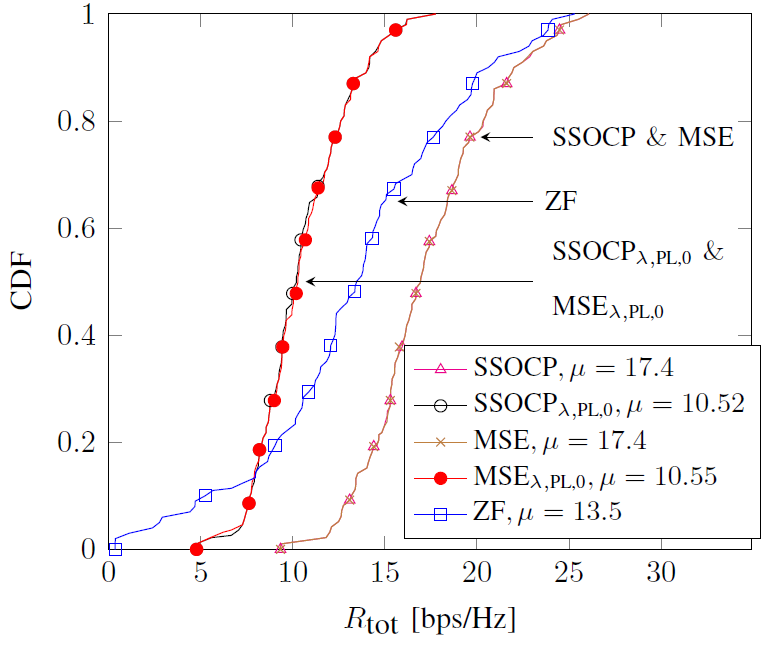}\protect\caption{\label{fig:Comparison-the-CDF-1-1}Comparison of the CDF of SSOCP
and MSE based precoders in terms of the weighted sum rate. It can
be observed that the curves SSOCP and MSE overlap. The $\mbox{SSOCP}_{\lambda,\text{PL,0}}$
and $\mbox{MSE}_{\lambda,\text{PL,0}}$ curves closely overlap, when
long term channel statistics is incorporated under limited CSI and
limited backhauling constraint. The cell-edge SNR is 15 dB and the
threshold is 3 dB. The $\mu$ values in the legend shows the average
value. }
\end{figure}
Fig.~\ref{fig:Average-Sum-Rate-1} captures the effect of increasing
the cell-edge SNR on the average weighted sum rate for a relative
threshold of 9~dB. In the case of having limited information and
long term channel statistics, $\mbox{SSOCP}_{\lambda,\text{PL,0}}$
can be useful compared to $\mbox{SSOCP}_{\text{0}}$. It is interesting
to note that the naive approach $\mbox{SSOCP}_{\text{PL,0}}$ only
performs well at low thresholds (not shown here), and the performance
deteriorates at high thresholds, and also with the increase in the
cell-edge SNR. 

Fig.~\ref{fig:Comparison-the-CDF-1-1} shows the cumulative distribution
function (CDF) of the weighted sum rate of the MSE and SSOCP based
precoder. The MSE approach achieves a performance similar to that
of the SSOCP. The MSE approach is very attractive due to the simple
sub-problems being solved in every iteration. However, it takes a
longer time for convergence due to the receiver updates.

\subsection{Effect of number of BS antennas}

\begin{figure}[tbh]
\begin{centering}
\includegraphics[width=0.4\paperwidth]{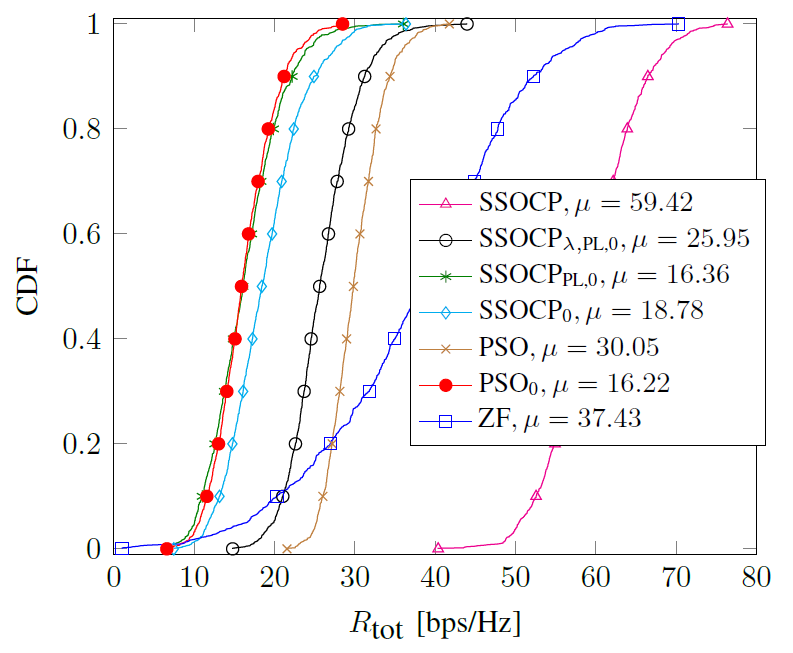}
\par\end{centering}

\protect\caption{\label{fig:With-cell-edge-SNR}With cell-edge SNR of 15 dB and a threshold
of 3 dB, $\left|\mathcal{B}\right|=3$ BSs with $N_{\text{T}}=3$
antennas each are serving 9 users. The $\mu$ values in the legend
shows the average value. }
\end{figure}
In this section, we investigate the effect of the number of BS antennas
on the performance of the precoders. Fig.~\ref{fig:With-cell-edge-SNR}
shows the CDF of the weighted sum rate where the number of antennas
is increased to $N_{\text{T}}=3$ at each of the 3 BSs, serving 9
users, resulting in a fully loaded system. For the non-SSOCP cases,
the per-antenna power constraint is applied just as in the case of
SSOCP. Apart from which, the whole precoding matrix is rescaled such
that at least one of the antennas is transmitting at maximum power.
Note that the per-antenna power constraint is more practical and that
the SSOCP is more capable of utilizing this to the fullest extent.
It can be observed that the proposed SSOCP outperforms all other precoding
algorithms. However, the naive approach $\mbox{SSOCP}_{\text{PL,0}}$
performs poorly. It is interesting to note that the SSOCP has consistent
cell-edge performance with steeper CDF curves compared to the ZF approach.

\begin{figure}[tbh]
\begin{centering}
\includegraphics[width=0.4\paperwidth]{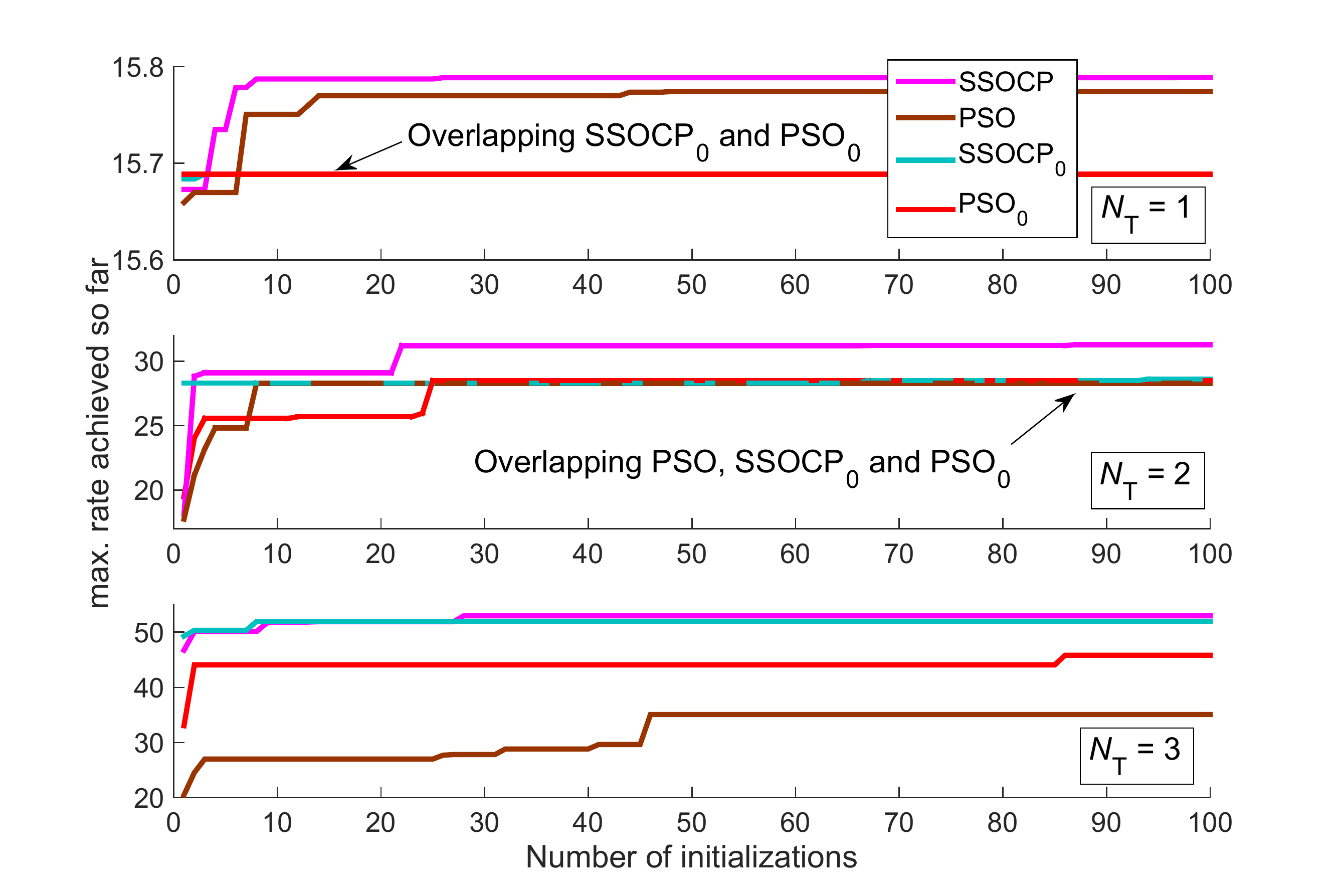}
\par\end{centering}

\protect\caption{\label{fig:A-comparison-of} A comparison of the performance of PSO
and SSOCP with the number of random initializations for a cell-edge
SNR of 15~dB and a relative threshold of 3~dB.}
\end{figure}
Fig.~\ref{fig:A-comparison-of} captures the maximum rate that is
achieved when designing the precoder at the central coordination node
based on the number of random initializations of the precoder. Recall
that increasing this number improves the chances of finding a solution
close to the global optimum \cite{TCJ08}. Each of the subplots in
Fig.~\ref{fig:A-comparison-of} show the impact on the achievable
rate of SSOCP and PSO with the increase in the problem size, due to
the increase in the number of antennas at each of the BSs in a fully
loaded system. With $N_{\text{T}}=1$, the 3 BSs serve $\left|\mathcal{U}\right|=3$
users. To keep the system fully loaded, we consider $\left|\mathcal{U}\right|=6$
when $N_{\text{T}}=2$ , and $\left|\mathcal{U}\right|=9$ when $N_{\text{T}}=3$.
For limited feedback, choosing $\mbox{MAXRETRIES}=5$ is good enough
when one considers the tradeoff between the number of initializations
and the achievable rate based on the available CSI at the central
coordination node. It is interesting to note that PSO performs closer
to SSOCP when the problem size is small, however, SSOCP outperforms
consistently with the increase in the problem size in terms of the
increase in the number of transmit antennas and the users. It should
be noted that the power allocation with PSO is merely a scaling of
the entire precoding matrix, as in \cite{ZD04}, and that with increased
problem size, PSO is unable to completely make use of the per-antenna
power constraint as in \eqref{eq:PerAntennaPwrConstr}. With $N_{\text{T }}=3$,
this behavior can be easily explained, and it can be concluded that
the PSO is unable to converge in polynomial time. Due to this we do
not consider PSO for any further analysis.

\subsection{Bounding the proposed SSOCP}

\begin{figure}[tbh]
\begin{centering}
\includegraphics[width=0.4\paperwidth]{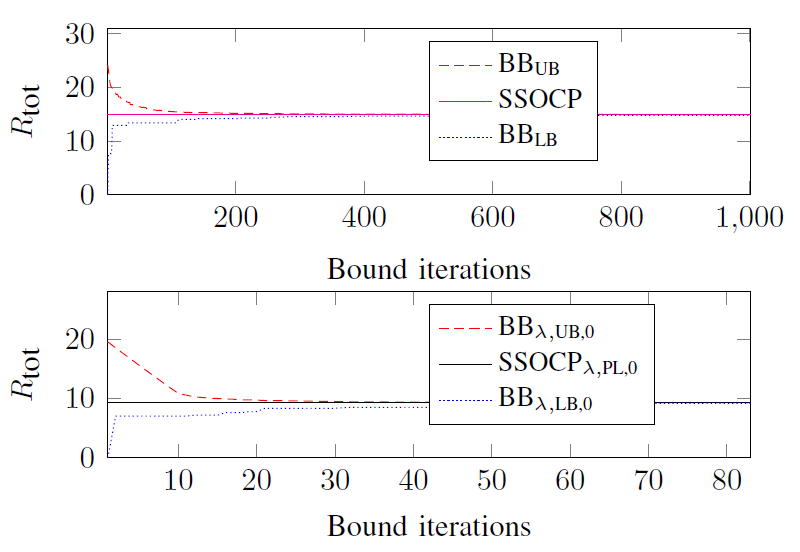}
\par\end{centering}

\centering{}\protect\caption{\label{fig:The-convergence-of-1}The convergence of the upper and
lower bounds from the branch and bound procedure used to benchmark
SSOCP and $\mbox{SSOCP}_{\lambda,\text{PL},0}$, for a given realization
of the aggregated channel matrix with full/limited feedback and backhauling
under $\left|\mathcal{B}\right|=3,N_{\text{T}}=1,\left|\mathcal{U}\right|=3$,
with cell-edge SNR of 15 dB and a relative threshold of 3 dB. The
$\mbox{MAXRETRIES}=5,20$ for SSOCP and $\mbox{SSOCP}_{\lambda,\text{PL},0}$,
respectively.}
\end{figure}
\begin{figure}[tbh]
\centering{}\includegraphics[width=0.4\paperwidth]{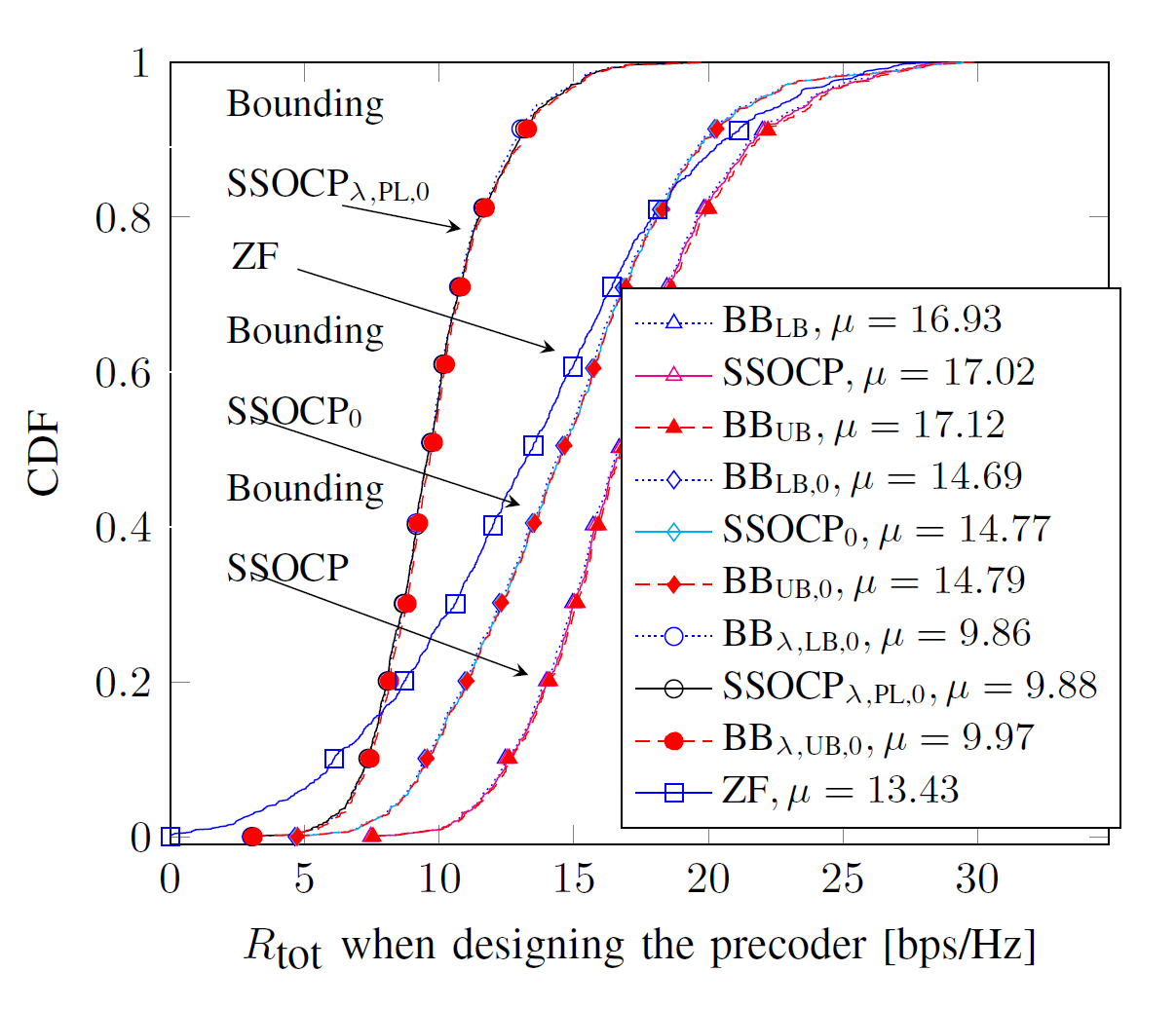}\protect\caption{\label{fig:Comparison-the-CDF-1}The plot shows the comparison of
the CDF of sum rate obtained when designing the precoder and their
corresponding bounds. The $\mu$ values in the legend shows the average
value. Here the MAXRETRIES = 20, in Algorithm~\ref{alg:SSOCP-algorithm-for}
for the $\mbox{SSOCP}_{\lambda,\text{PL},0}$, whereas a default value
of MAXRETRIES = 5 results in a performance slightly lower than the
lower bound.}
\end{figure}
In this section, we show that the accuracy of the solution obtained
with SSOCP is tightly bounded. Here, we restrict our simulations to
the SSOCP based approach, as we have already observed in Fig.~\ref{fig:Comparison-the-CDF-1-1}
that the MSE and SSOCP have similar performance. Fig.~\ref{fig:The-convergence-of-1}
shows the convergence of the branch and bound algorithm under full/limited
feedback and backhauling conditions. The y-axis captures the weighted
sum rate obtained when designing the precoder at the central coordination
node. Notice that the convergence is slow with $\left(\mbox{BB}_{\text{UB}},\mbox{BB}_{\text{LB}}\right)$
in the topmost subplot, when compared to the convergence behavior
of the other bounds. This is due to the size of the problem. Also
notice that the numerical bounds tightly characterize the proposed
SSOCP algorithm for precoder design. It should be mentioned that the
bounds can be tightened depending on step~\ref{-Set-tolerence} in
Algorithm~\ref{alg:The-branch-and}. The branch and bound technique
is extremely slow in the CVX \cite{GB11} environment even with the
bisection method (Algorithm~\ref{alg:Subroutine:-Bisection-method})
being applied to improve the lower bound.  Likewise the CDF of the
bounds are captured in Fig.~\ref{fig:Comparison-the-CDF-1} for full
and limited feedback and backhauling, with and without the use of
long term statistics. It can be observed that the BB technique tightly
bounds the proposed SSOCP. Note that when limited information is considered,
the curves are those that were obtained during the precoder design
at the coordination node. It can be observed that when the long term
statistics are included, the precoder rate is more pessimistic. However,
they perform better during actual transmission when the complete channel
is considered as seen in Fig.~\ref{fig:With-PL+SF}.

\section{Conclusions\label{sec:Conclusions}}

In this work, we have incorporated the long term channel statistics
in the SSOCP algorithm to efficiently solve the precoder design when
there is limited channel state information available at the central
coordination node. Efficient backhauling is achieved, in a sense that
the number of precoding weights generated for the active links is
equal to the number of coefficients of the channel state information
correspondingly available for the active links at the central coordination
node. The above goals are accomplished with the objective of maximizing
the weighted sum rate when jointly transmitting to a group of cell-edge
users. The efficiency of the solution obtained with the proposed SSOCP
algorithm is verified by the tight upper and the lower bounds. The
performance of the precoder is also studied for various thresholds,
cell-edge SNRs and with the increase in the problem size. Alternatively,
we also derived the weighted MSE approach for the precoder design
that incorporates the long term channel statistics when there is limited
information, and it was shown to achieve the performance of SSOCP.

\appendices{}

\section{The receive variance with limited CSI and long term channel statistics\label{AppRxCov}}

To obtain \eqref{eq:RxCovWithLambda}, consider \eqref{eq:RxCov}
\[
\begin{split}\widetilde{c}_{u} & =\underset{\forall i}{\sum}\underset{b\in\mathcal{B}_{u}}{\sum}\mathbf{h}_{b,u}\mathbf{w}_{b,i}\left(\mathbf{h}_{b,u}\mathbf{w}_{b,i}\right)^{H}+N_{0}\\
 & =\underset{b\in\mathcal{B}_{u}}{\sum}\mathbf{h}_{b,u}\mathbf{w}_{b,u}\left(\mathbf{h}_{b,u}\mathbf{w}_{b,u}\right)^{H}+\underset{i\neq u}{\sum}\underset{b\in\mathcal{B}_{u}}{\sum}\mathbf{h}_{b,u}\mathbf{w}_{b,i}\left(\mathbf{h}_{b,u}\mathbf{w}_{b,i}\right)^{H}+N_{0}\\
 & =\left|\underset{b\in\mathcal{B}_{u}}{\sum}\mathbf{h}_{b,u}\mathbf{w}_{b,u}\right|^{2}+\underset{i\neq u}{\sum}\left|\underset{b\in\mathcal{B}_{i}}{\sum}\mathbf{h}_{b,u}\mathbf{w}_{b,i}\right|^{2}+N_{0}.
\end{split}
\]
With limited information, we consider the expected value of the inactive
links in the interference terms as
\[
\begin{split}c_{u} & =N_{0}+\left|\underset{b\in\mathcal{B}_{u}}{\sum}\mathbf{h}_{b,u}\mathbf{w}_{b,u}\right|^{2}+E_{\overline{h}}\underset{i\neq u}{\sum}\left\{ \left|\underset{b\in\mathcal{B}_{i}\cap\mathcal{B}_{u}}{\sum}\mathbf{h}_{b,u}\mathbf{w}_{b,i}+\underset{b\in\mathcal{B}_{i}\backslash\mathcal{B}_{u}}{\sum}\overline{\mathbf{h}}_{b,u}\mathbf{w}_{b,i}\right|^{2}\right\} \\
 & \leq N_{0}+\left|\underset{b\in\mathcal{B}_{u}}{\sum}\mathbf{h}_{b,u}\mathbf{w}_{b,u}\right|^{2}+\underset{i\neq u}{\sum}\left\{ \left|\underset{b\in\mathcal{B}_{i}\cap\mathcal{B}_{u}}{\sum}\mathbf{h}_{b,u}\mathbf{w}_{b,i}\right|^{2}+|\mathcal{B}_{i}\backslash\mathcal{B}_{u}|\underset{b\in\mathcal{B}_{i}\backslash\mathcal{B}_{u}}{\sum}E_{\overline{h}}\left|\overline{\mathbf{h}}_{b,u}\mathbf{w}_{b,i}\right|^{2}\right\} \\
 & =N_{0}+\left|\underset{b\in\mathcal{B}_{u}}{\sum}\mathbf{h}_{b,u}\mathbf{w}_{b,u}\right|^{2}+\underset{i\neq u}{\sum}\left\{ \left|\underset{b\in\mathcal{B}_{i}\cap\mathcal{B}_{u}}{\sum}\mathbf{h}_{b,u}\mathbf{w}_{b,i}\right|^{2}+|\mathcal{B}_{i}\backslash\mathcal{B}_{u}|\underset{b\in\mathcal{B}_{i}\backslash\mathcal{B}_{u}}{\sum}\lambda_{b,u}^{2}||\mathbf{w}_{b,i}||_{2}^{2}\right\} .
\end{split}
\]
Steps similar to \eqref{eq:StartNewSINR}-\eqref{eq:SINR_PL} are
applied here.

\section{Cookbook version of the branch and bound that incorporates the long
term channel statistics\label{AppCookboockBB}}

For completeness of Section~\ref{sub:Branch-and-Bound}, a cookbook
version of the branch and bound based on \cite{JWC+12} is provided
in Algorithm~\ref{alg:The-branch-and}, with emphasis on including
the long term channel statistics into the precoder design, where the
SINRs are checked for feasibility in Algorithm~\ref{alg:-Check-if}.

\begin{algorithm}[H]
\begin{centering}
\protect\caption{\label{alg:The-branch-and}The branch and bound algorithm for bounding
with limited information, and applying bisection method to improve
$\gamma_{\text{max}}$ and return the upper bound $BB_{\text{UB}}$
and the lower bound $BB_{\text{LB}}$ of the objective in \eqref{eq:sumRate}.}

\par\end{centering}

\small{

\begin{algorithmic}[1]

\STATE \label{-Set-tolerence}Set tolerance $\epsilon=0.1$, $maxIter=100$ 

\STATE Set $\mathcal{Q}_{\text{curr}}=\mathcal{Q}_{\text{init}}$

\STATE Algorithm~\ref{alg:Subroutine:-Bisection-method}: Bisection
method to limit $\gamma_{\text{max}}$ of $\mathcal{Q}_{\text{curr}}$

\STATE Algorithm~\ref{alg:-Check-if}: Check if $\gamma_{\text{min}}$
of $\mathcal{Q}_{\text{curr}}$ is feasible

\STATE Algorithm~\ref{alg:-Update-the}: Update $BB{}_{\text{UB}}$,
$BB{}_{\text{LB}}$ based on the above feasibility

\STATE Accumulate hyperrectangles and the corresponding bounds:\\
 $\mathcal{A}=\left\{ \left(\mathcal{Q}_{\text{curr}},B_{\text{UB}}(\mathcal{Q}_{\text{curr}}),B{}_{\text{LB}}(\mathcal{Q}_{\text{curr}})\right)\right\} $

\WHILE { $BB_{\text{UB}}-BB{}_{\text{LB}}>\epsilon$ AND $maxIter$}

\FOR { $a\in\mathcal{A}$ }

\IF{ $BB{}_{\text{LB}}=a.B_{\text{LB}}$ }

\STATE $\mathcal{Q}_{\text{curr}}=a.\mathcal{Q}$

\STATE $a_{\text{curr}}=a$

\STATE \textbf{break}

\ENDIF

\ENDFOR

// Branching

\STATE Split the longest edge of hyperrectangle $\mathcal{Q}_{\text{curr}}$
into $\mathcal{Q}_{1}$ and $\mathcal{Q}_{2}$

\FOR { $i=1,2$ }

\STATE Algorithm~\ref{alg:Subroutine:-Bisection-method}: Bisection
method to limit $\gamma_{\text{max}}$ of $\mathcal{Q}_{i}$

\STATE Algorithm~\ref{alg:-Check-if}: Check if $\gamma_{\text{min}}$
of $\mathcal{Q}_{i}$ is feasible \label{-Algorithm:-If}

\STATE Algorithm~\ref{alg:-Update-the}: Update $B_{\text{UB}}(\mathcal{Q}_{i})$
and $B_{\text{LB}}(\mathcal{Q}_{i})$ 

\ENDFOR

\STATE Remove $\left\{ a_{\text{curr}}\right\} $ from $\mathcal{A}$

\STATE Update $\mathcal{A}=\mathcal{A}\cup$\\
$\left\{ \left(\mathcal{Q}_{1},B_{\text{UB}}(\mathcal{Q}_{1}),B{}_{\text{LB}}(\mathcal{Q}_{1})\right),\left(\mathcal{Q}_{2},B_{\text{UB}}(\mathcal{Q}_{2}),B{}_{\text{LB}}(\mathcal{Q}_{2})\right)\right\} $

// Bounding

\STATE $BB{}_{\text{UB}}=\underset{a\in\mathcal{A}}{\mbox{min}}\left(a.B_{\text{UB}}\right)$

\STATE $BB{}_{\text{LB}}=\underset{a\in\mathcal{A}}{\mbox{min}}\left(a.B_{\text{LB}}\right)$ 

\STATE $maxIter=maxIter-1$

\ENDWHILE

\RETURN { $BB_{\text{UB}},BB_{\text{LB}},\mathbf{W}^{\star}$ from
step~\ref{-Algorithm:-If}}

\end{algorithmic}

}
\end{algorithm}

\begin{algorithm}[H]
\begin{centering}
\protect\caption{\label{alg:Subroutine:-Bisection-method}Bisection method to improve
$\gamma_{\text{max}}$ for the lower bound.}

\par\end{centering}

\small{

\begin{algorithmic}[1]

\STATE Set tolerance $\epsilon=0.01$

\FOR { Each user }\label{-Each-UE}

\STATE $\mathbf{a}=\gamma_{\text{min}}+\left(\gamma_{u,\text{max}}-\gamma_{u,\text{min}}\right)\cdot\mathbf{e}_{u}$
where $\mathbf{e}_{u}$ is the standard basis vector

\STATE Set $\mathbf{b}_{\text{lower}}=\gamma_{\text{min}}$ and $\mathbf{b}_{\text{upper}}=\mathbf{a}$

\STATE Algorithm~\ref{alg:-Check-if}: Check if $\mathbf{a}$ is
feasible

\IF{ feasible }

\STATE $\gamma_{u,\text{max}}^{\star}=\mathbf{a}$ 

\STATE \textbf{continue} step~\ref{-Each-UE} // Avoid unnecessary
bisection steps below

\ENDIF

\WHILE { $||\mathbf{b}_{\text{upper}}-\mathbf{b}_{\text{lower}}||_{2}>\epsilon$
}

\STATE $\mathbf{t}=\left(\mathbf{b}_{\text{lower}}+\mathbf{b}_{\text{upper}}\right)/2$

\STATE Algorithm~\ref{alg:-Check-if}: Check if $\mathbf{t}$ is
feasible

\IF{ feasible }

\STATE $\mathbf{b}_{\text{lower}}=\mathbf{t}$

\ELSE 

\STATE $\mathbf{b}_{\text{upper}}=\mathbf{t}$

\ENDIF

\ENDWHILE

\STATE $\mathbf{\gamma_{u,\text{max}}^{\star}}=\mathbf{b}_{\text{upper}}$

\ENDFOR

\RETURN { $\gamma_{\text{max}}^{\star}=\left[\gamma_{1,\text{max}}^{\star},\ldots,\gamma_{u,\text{max}}^{\star},\ldots,\gamma_{\left|\mathcal{U}\right|,\text{max}}^{\star}\right]$}

\end{algorithmic}

}
\end{algorithm}

\begin{algorithm}[H]
\begin{centering}
\protect\caption{\label{alg:-Update-the} Update the bounds based on a given hyperrectangle
$\mathcal{Q}_{\text{given}}$ and its feasibility.}

\par\end{centering}

\small{

\begin{algorithmic}[1]

\IF{ feasible }

\STATE Set $\gamma_{u}=\gamma_{u,\text{min}}$

\STATE Evaluate \eqref{eq:sumRate}, $B_{\text{U}}(\mathcal{Q}_{\text{given}})=R_{\text{tot}}$

\STATE Set $\gamma_{u}=\gamma_{u,\text{max}}$

\STATE Evaluate \eqref{eq:sumRate}, $B_{\text{L}}(\mathcal{Q}_{\text{given}})=R_{\text{tot}}$

\ELSE 

\STATE set $B_{\text{U}}(\mathcal{Q}_{\text{given}})=0$ and $B_{\text{L}}(\mathcal{Q}_{\text{given}})=0$ 

\ENDIF

\RETURN { $B_{\text{U}}(\mathcal{Q}_{\text{given}}),B_{\text{L}}(\mathcal{Q}_{\text{given}})$}

\end{algorithmic}

}
\end{algorithm}

\section*{Acknowledgment}

We would like to thank the colleagues at the Centre of Wireless Communications,
University of Oulu, for the friendly, helpful, and interesting discussions.
Also, we would like to thank the members of the VR project meetings
with Uppsala University and Karlstad University for their comments.
Part of this work has been performed in the framework of the FP7 project
ICT-317669 METIS, which is partly funded by the European Union. This
work is also supported by the Swedish Research Council VR under the
project 621-2009-4555 Dynamic Multipoint Wireless Transmission. Some
computations were performed on $\mbox{C}^{3}\mbox{SE}$.

\end{document}